\documentclass{article}

\usepackage{microtype}
\usepackage{caption}
\usepackage{subcaption}
\usepackage{graphicx}
\usepackage{booktabs} 
\usepackage{mathtools}

\allowdisplaybreaks
\usepackage{hyperref}

\usepackage[accepted]{icml2024}

\usepackage{amsmath}
\usepackage{amssymb}
\usepackage{mathtools}
\usepackage{amsthm}

\usepackage[capitalize,noabbrev]{cleveref}

\theoremstyle{plain}
\newtheorem{theorem}{Theorem}[section]

\newtheorem{lemma}[theorem]{Lemma}

\theoremstyle{definition}
\newtheorem{definition}[theorem]{Definition}

\theoremstyle{remark}

\DeclareMathOperator*{\argmax}{arg\,max}

\usepackage[textsize=tiny]{todonotes}

\icmltitlerunning{Rationality of Learning Algorithms in Repeated Normal-Form Games}

\begin{document}

\twocolumn[
\icmltitle{Rationality of Learning Algorithms in Repeated Normal-Form Games}



\icmlsetsymbol{equal}{*}

\begin{icmlauthorlist}
\icmlauthor{Shivam Bajaj}{yyy}
\icmlauthor{Pranoy Das}{yyy}
\icmlauthor{Yevgeniy Vorobeychik}{comp}
\icmlauthor{Vijay Gupta}{yyy}
\end{icmlauthorlist}

\icmlaffiliation{yyy}{Department of Electrical and Computer Engineering, Purdue University, West Lafayette, Indiana, USA}
\icmlaffiliation{comp}{Department of Computer Science and Engineering, Washington University in St. Luis, St. Luis, Missouri, USA}

\icmlcorrespondingauthor{Shivam Bajaj}{bajaj41@purdue.edu}


\vskip 0.3in
]




\begin{abstract}
Many learning algorithms are known to converge to an equilibrium for specific classes of games if the same learning algorithm is adopted by all agents. However, when the agents are self-interested, a natural question is whether agents have a strong incentive to adopt an alternative learning algorithm that yields them greater individual utility.
We capture such incentives as an algorithm's \emph{rationality ratio}, which is the ratio of the highest payoff an agent can obtain by deviating from a learning algorithm to its payoff from following it.
We define a learning algorithm to be \emph{$c$-rational} if its rationality ratio is at most $c$ irrespective of the game.
We first establish that popular learning algorithms such as fictitious play and regret matching are not $c$-rational for any constant $c\geq 1$. We then propose and analyze two algorithms that are provably $1$-rational under mild assumptions, and have the same properties as (a generalized version of) fictitious play and regret matching, respectively, if all agents follow them. Finally, we show that if an assumption of perfect monitoring is not satisfied, there are games for which $c$-rational algorithms do not exist, and illustrate our results with numerical case studies.
\end{abstract}

\section{Introduction}
Learning in a multi-agent framework with cooperative, non-cooperative, or mixed agents \cite{jacq2020foolproof,nguyen2019non,lowe2017multi,zhang2021mfvfd}  has been expanding rapidly with proposed applications including cyber-physical systems \cite{adler2002cooperative,bashendy2023intrusion}, finance \cite{lee2007multiagent}, sensor networks \cite{choi2009distributed}, and robotics \cite{8807386,orr2023multi}. 
A popular framework in this area is to model for the interactions among self-interested and rational agents as a  repeated or stochastic game ~\cite{yang2020overview,hu1998multiagent,littman1994markov,bowling2002multiagent,zhang2021multi}. Given that computation of an equilibrium of the underlying game is often not computationally feasible, a fundamental line of research in this model is to identify simple learning procedures that, if adopted by all agents (i.e., in \emph{self-play}), converge to an equilibrium \cite{monderer1996fictitious,leslie2006generalised,sayin2022fictitious,baudin2022fictitious,jacq2020foolproof,chasnov2019convergence}. 
A long line of work has identified many algorithms, including fictitious play and regret-matching, that converge to an equilibrium for specified classes of games \cite{fudenberg1998theory,robinson1951iterative,hart2000simple}.
This was subsequently followed up by a literature that aimed to develop learning algorithms that, additionally, satisfy other desirable properties, such as converging to a policy that is optimal if facing a stationary opponent~\cite{bowling2002multiagent,conitzer2007awesome}.
A closely related line of work considered \emph{rational learning} in which agents use Bayesian belief updates, with the classic result by \citet{kalai1993rational} showing convergence of such learning to a Nash equilibrium of the corresponding extensive form game.
However, this result requires that the distribution over histories induced by the actual strategies is continuous with respect to the distribution induced by an agent's beliefs \cite{shoham2008multiagent}. This condition has been argued to be restrictive \cite{nachbar2001bayesian} and there exists games for which such an algorithm does not converge to Nash equilibrium \cite{foster2001impossibility}. 

Even in the cases of equilibrium convergence in self-play, however, the learning algorithms themselves need not constitute equilibrium behavior of the corresponding \emph{dynamic} (e.g., stochastic, or repeated) game.
Indeed, since the nature of the very games under consideration is non-cooperative, insofar as there are strong incentives for individual agents to deviate from learning algorithms, we would expect them to do so~\cite{vundurthy2023intelligent}.

In their classic work, \citet{brafman2002efficient} address this issue by introducing the notion of a \emph{learning equilibrium} which requires that the learning algorithms are {\em rational}, in the sense that self-play is a symmetric equilibrium of the dynamic game, and proposing specific learning algorithms that indeed constitute an equilibrium in self-play.
However, they assumed a uniform bound on game payoffs, and did not connect this equilibrium notion to more conventional  learning approaches in games, such as fictitious play or regret matching.
A natural question remains whether such conventional approaches are already nearly in equilibrium in self-play, and if not, whether we can induce this property in a way that is outcome-equivalent to these algorithms in self-play (thereby satisfying whatever other desirable properties these algorithms possess, such as no-regret learning~\cite{greenwald2006bounds,hart2000simple}).



Designing learning algorithms that are rational (i.e., constitute a near-equilibrium) in self-play is challenging, since such algorithms may not even exist for certain classes of games. 
We will see later in Theorem \ref{thm:imperfect_games} that this is indeed possible. In such cases, a quantifiable metric that characterizes how much an agent may benefit by deviating from its learning algorithm is required to allow the system designer to compare various learning algorithms and choose one appropriately.

To this end, we consider a two-agent repeated game framework under an assumption of perfect monitoring and introduce \emph{rationality ratio}, defined as the ratio of the most an agent can obtain by deviating from a learning algorithm to their payoff from following it. A learning algorithm is $c$-rational if its rationality ratio is no more than $c$ in the worst-case. Finally, a learning algorithm is said to be \emph{perfectly rational} if $c=1$.
We first establish that classic learning algorithms such as fictitious play and regret matching algorithms are not $c$-rational for any given constant $c$. We then provide two algorithms which, under certain conditions, are provably perfectly rational. We also establish that there exists games for which $c$-rational algorithms do not exist if the assumption of perfect monitoring does not hold.

 
Apart from the line of literature discussed above that started from~\cite{brafman2002efficient,bowling2002multiagent,conitzer2007awesome}, other works closely relevant to ours include~\cite{jacq2020foolproof,digiovanni2022balancing}. However, in contrast to these works, we consider non-cooperative agents and do not impose any assumption on the strategies followed by the agents upon deviating from the specified algorithm.



This work is organized as follows. Section \ref{sec:model} reviews some basic concepts and formally describes the model as well as the rationality ratio. Section \ref{sec:irrationality} establishes that the rationality ratio of fictitious play and regret matching algorithms is unbounded in the worst-case. Section \ref{sec:Algorithms} presents two algorithms that are provably perfectly rational. Finally, Section \ref{sec:numerics} provides numerical insights into our algorithms and Section \ref{sec:conclusion} outlines some directions for future work.

\section{Model and Definitions}\label{sec:model}
We first review some basic concepts in game theory and present the model considered in this work. A \emph{game} describes an interaction among a set of agents, wherein the rules of the game describe the order of moves by the agents, the information available to each agent, and the final outcome for each agent.
Formally, a stage game  is  defined as follows.

\begin{definition}[Stage Game]
   A two agent stage game is a tuple $\mathcal{G}=(A_1,A_2,\mathcal{R}_1,\mathcal{R}_2)$, where $A_i$, $i\in\{1,2\}$, denotes a finite set of actions available to agent $i$ and $\mathcal{R}_i:A_1\times A_2\to\mathbb{R}$ is a payoff function for agent $i$. A \emph{strategy} $\pi_i$ for agent $i$ is a probability distribution over its action set $A_i$. A \emph{pure strategy} is a strategy in which the probability of selecting a particular action is set to one. A \emph{mixed strategy} is a probability distribution over pure strategies.
\end{definition}
Following standard notation, when referring to an agent $i$, we will refer to the other agent as the agent $-i$. Further, for simplicity, we will assume that $\mathcal{R}_i>0$.

\begin{figure}[t]
\centering
  \begin{subfigure}[b]{0.49\columnwidth}
  \centering
    \includegraphics[width=\linewidth]{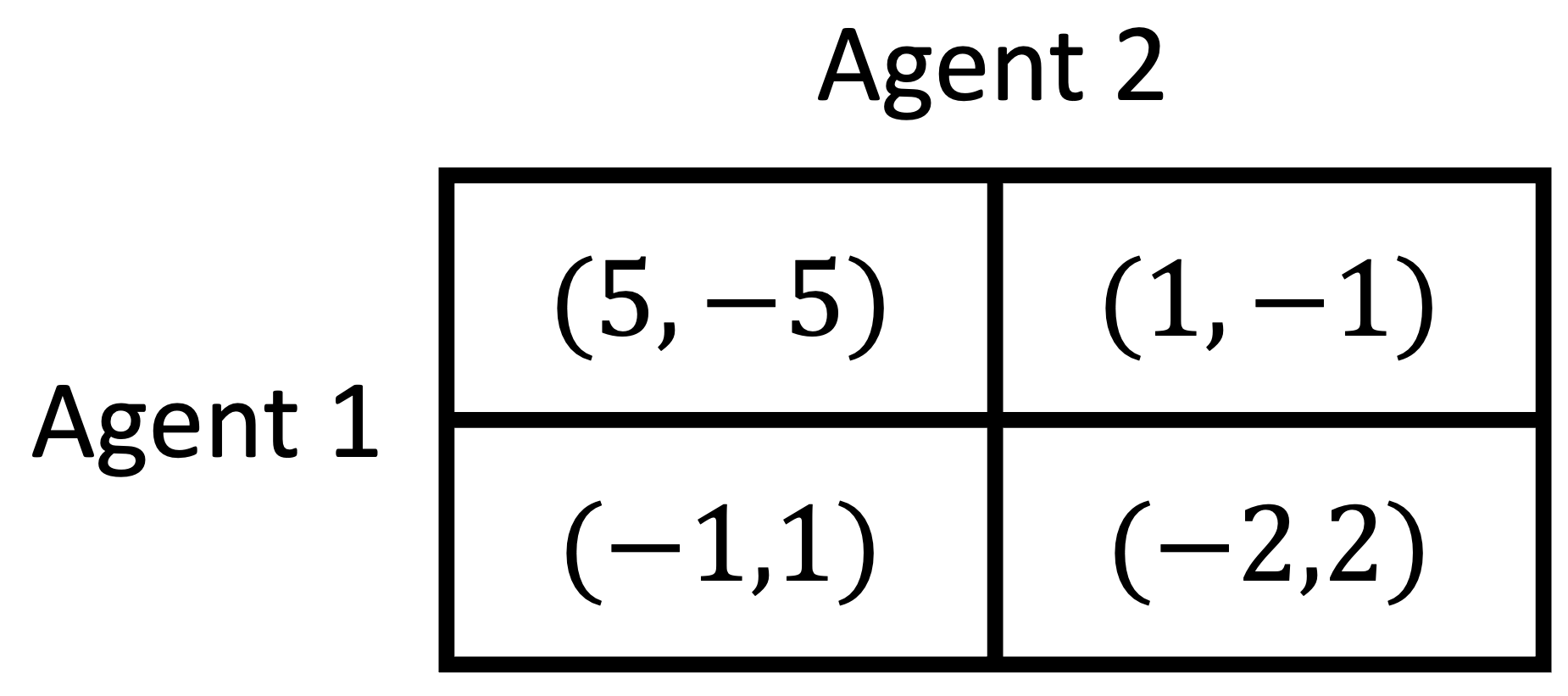}
    \caption{A zero-sum stage game.}
    \label{fig:eg_1}
  \end{subfigure}
  \hfill 
  \begin{subfigure}[b]{0.49\columnwidth}
  \centering
    \includegraphics[width=\linewidth]{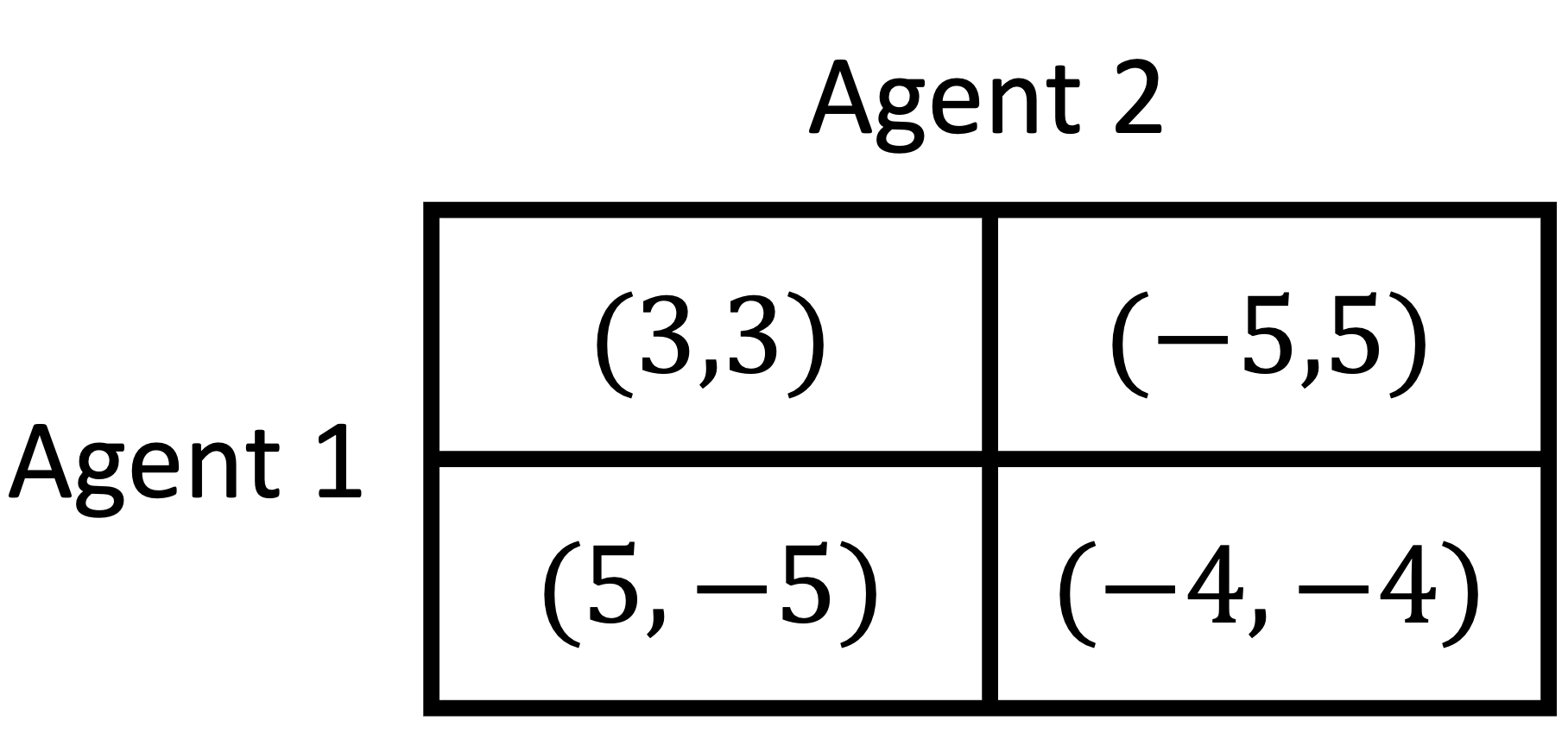}
    \caption{A general-sum stage game.}
    \label{fig:eg_2}
  \end{subfigure}
  \caption{Illustration of two player stage games.}
  \label{fig:examples_games}
\end{figure}

We will focus on the class of stage games for two agents that can be described using a bi-matrix whose rows (resp. columns) correspond to the possible actions of the first (resp. second) agent. The $(j,k)$ entry of the bi-matrix contains a pair of values denoting the payoffs to each agent when agent 1 (resp. agent 2) plays action $j$ (resp. $k$). The payoffs corresponding to the entry $(j,k)$ in the bi-matrix is denoted as $(r^1_{j,k},r^2_{j,k})$. When referring to the payoffs of the agents separately, we will use $R^1$ and $R^2$ to denote the payoff matrix of agent $1$ and agent $2$, respectively. Thus, $r^1_{j,k}$ (resp. $r^2_{j,k}$) corresponds to the $(j,k)$ entry of matrix $R^1$ (resp. $R^2$).
Figure \ref{fig:examples_games} illustrates two bi-matrix stage games. The first is a \emph{zero-sum} stage game in which the sum of the payoffs of the agents is $0$ and the second is a general sum stage game, in which the agents are neither pure competitors nor they have identical interests.

The agents' strategies are in equilibrium if no agent unilaterally benefits by changing its strategy, given the other agent's strategy. One such equilibrium is defined as follows. 
\begin{definition}[Nash Equilibrium]
    A  strategy profile $(\pi_1^*,\pi_2^*)$ is a Nash equilibrium if
    \begin{align*}
        &\mathcal{R}_1(\pi_1^*,\pi_2^*)\geq  \mathcal{R}_2(\pi_1,\pi_2^*), \forall \pi_1\neq \pi_1^*,\\
        &\mathcal{R}_2(\pi_1^*,\pi_2^*)\geq  \mathcal{R}_2(\pi_1^*,\pi_2), \forall \pi_2\neq \pi_2^*.
    \end{align*}
\end{definition}
The pair $(\mathcal{R}_1(\pi_1^*,\pi_2^*),\mathcal{R}_2(\pi_1^*,\pi_2^*))$ is known as the Nash outcome of the stage game.

Unfortunately, calculation of Nash equilibria is known to be computational difficult. One of the commonly used model for how agents can learn these equilibria is that of a \emph{repeated game}. In a repeated game, the agents play a given stage game repeatedly. At each iteration or time step of the stage game, the agents observe their (and possibly the other agent's) rewards and actions from all previous iterations of the stage game. Based on the observations made by the agent, the agent may update its strategy.
An \emph{algorithm} for agent $i$ is a mapping from the set of possible histories $H$ to an action $a_i\in A_i$ at every iteration.

A fundamental component of a learning algorithm is to specify what information about the stage game is known to the agents. A payoff matrix $R^i$ is said to be \emph{completely known} to agent $i$ if agent $i$ has the information of all entries of $R^i$. Similarly, a row $j$ (resp. column $k$) of $R^i$ is said to be \emph{completely known} to agent $i$ if agent $i$ has the information of the all of the entries of $j$th row (resp. $k$th column) of $R^i$.  In our work, we assume that, in the first iteration, the agents do not have any information about their own as well as the other agent's payoff matrices. Instead, we assume  a \emph{perfect monitoring} setting in which, at each iteration or time $t$ of the stage game, an agent can observe the actions as well as the payoffs received by both agents. 

Given a repeated game formed through iterations of the stage game $\mathcal{G}$, the \emph{value} for agent $1$ is defined as $U_1(\mathcal{G},\mathcal{A}_{1},\mathcal{A}_{2})=\lim\inf_{T\to \infty} \mathbb{E}[\tfrac{1}{T}\sum_{t=0}^T \mathcal{R}_{1,t}]$, where $\mathcal{A}_{1}$ and $\mathcal{A}_{2}$ denote the learning algorithm followed by agent $1$ and agent $2$, respectively, and $\mathcal{R}_{1,t}$ (resp. $\mathcal{R}_{2,t}$) is the payoff received by agent $1$ (resp. agent $2$) at time or iteration $t$ of the stage game.
In what follows, to avoid notational clutter, we will drop the dependence on $\mathcal{G}$ from the notation of $U_i(\mathcal{G},\mathcal{A}_{1},\mathcal{A}_{2})$, $i\in\{1,2\}$ and write the term as $U_i(\mathcal{A}_{1},\mathcal{A}_{2})$. Note that $U_i(\mathcal{A}_1,\mathcal{A}_2)>0$ given our assumption that $\mathcal{R}_i>0$ for a stage game $\mathcal{G}$.

Traditionally, the learning algorithms followed by the agents are assumed to be prescribed a priori and further often fixed to the same choice for both the agents. Instead, we say that agent $i$ \emph{deviates} from a prescribed learning algorithm $\mathcal{A}$ if agent $i$ selects its actions according to any other algorithm $\mathcal{A}'$ in at least one interval of times or iterations $[t_1,t_2]$ for any $t_2\geq t_1\geq 0$. Note that there may exist more than one interval in which an agent $i$ deviates.

We now define a quantity that characterizes how much an agent $i$ gains by deviating from a learning algorithm $\mathcal{A}$. 

\begin{definition}[Rationality Ratio]
    Suppose agent $1$ deviates from a self-play algorithm $\mathcal{A}$ to any algorithm $\mathcal{A}'$. Then, the \emph{rationality ratio} of algorithm $\mathcal{A}$ is defined as
    \begin{equation}\label{eq:security_ratio}   s(\mathcal{A}',\mathcal{A})\coloneqq\frac{U_1(\mathcal{A}',\mathcal{A})}{U_1(\mathcal{A},\mathcal{A})}.
    \end{equation}
    Further, we say that algorithm $\mathcal{A}$ is $c$-\emph{rational} if
    \begin{equation}\label{eq:secure_algo}
    \sup_{\mathcal{G},\mathcal{A}'} s(\mathcal{A}',\mathcal{A}) \leq c,
\end{equation}
for a constant $c$. Finally, an algorithm $\mathcal{A}$ is \emph{perfectly rational} if $c=1$.
\end{definition}
The definition when agent $2$ deviates is analogous.

When $c=1$, then it means that there is no benefit for an agent to deviate from algorithm $\mathcal{A}$. 


We now review two classic self-play learning algorithms, namely fictitious play and regret matching, that are are known to converge to the Nash and correlated equilibria, respectively, for a wide class of games. We refer the reader to the long line of literature starting from~\cite{hart2000simple,brown1951iterative} for more details on these algorithms.

In fictitious play, at each time $t$, each agent $i$ selects an action that corresponds to the \emph{best-response} to the empirical frequency of the actions played by agent $-i$. Formally,
\begin{equation}\label{eq:FP}
    BR_i^{\text{Fict}}(\hat{\textbf{a}}_{-i}(t-1))\coloneqq \argmax_{a\in A_i} \mathcal{R}_i(a,\hat{\textbf{a}}_{-i}(t-1)),
\end{equation}
where $\hat{\textbf{a}}_{-i}(t)$ denotes the vector of empirical frequencies of actions $a_{-i}\in A_{-i}$ played until time $t$.

To describe the regret matching algorithm, we first define the \emph{instantaneous} regret of agent $i$ at time $t$ for action $a\in A_i$ as $\delta_{i}^t(a) \coloneqq \mathcal{R}_i(a,a_{-i}(t)) - \mathcal{R}_i(a_i(t),a_{-i}(t))$.
Further, we define the average regret of agent $i$ for action $a\in A_i$ at time $T$ as
\begin{align*}
    \delta^{\text{avg}}_{T,i}(a) \coloneqq \frac{1}{T}\sum_{t=1}^T \delta_{i}^t(a).
\end{align*}
Let $\delta^{\text{avg}}_+(a)$ denote the positive regret defined as $\delta^{\text{avg}}_+(a) \coloneqq \max\{0,\delta^{\text{avg}}_{T,i}(a)\}$. The regret-matching algorithm requires agent $i$ to pick an action with probability proportional to the positive regret on that action. Specifically, 
at each time $t+1$, agent $i$ selects action $a\in A_i$ with probabilities
\begin{equation}\label{eq:rm_prob}
    p_{t+1}^i(a) = 
    \begin{cases}
        \frac{\delta^{\text{avg}}_+(a)}{\sum_{a'\in A_i}\delta^{\text{avg}}_+(a')}, \text{ if } \sum_{a'\in A_i}\delta^{\text{avg}}_+(a')>0,\\
        \frac{1}{|A_i|}, \text{ otherwise,}
    \end{cases}
\end{equation}
where $|\cdot|$ denotes the cardinality of a set.

\section{Irrationality of existing self-play algorithms}\label{sec:irrationality}

The following results provide a discouraging result -- some commonly used learning algorithms are not $c$-rational for any given constant $c$. The proofs of all of the results are contained in the Appendix. 
\begin{theorem}\label{thm:BR_not_secure}
Fictitious play algorithm is not $c$-rational for any given constant $c\geq 1$.
\end{theorem}




\begin{theorem}\label{thm:RM_not_secure}
    Regret matching algorithm is not $c$-rational for any given constant $c\geq 1$.
\end{theorem}

Given these negative results, it is natural to ask whether {\em any} 
$c$-rational learning algorithms exist. Fortunately, the answer is in the affirmative. In the next section, we will design two algorithms that are provably perfectly rational (or equivalently $1$-rational).

\section{Rational Learning Algorithms}\label{sec:Algorithms}
We will now present two new learning algorithms and establish that they are perfectly rational. The two algorithms, summarized in Algorithm~\ref{alg:Sec-FP} and Algorithm~\ref{alg:Sec-RM}, build upon the classical fictitious play and regret matching algorithms, respectively. 
%
The idea behind these algorithms is to specify two strategies. The first is a strategy that is followed by the agent $i$ as long as agent $-i$ follows the same strategy. However, if agent $-i$ is detected to deviate from this strategy, agent $i$ switches to a prescribed {\em punishment} strategy. 
Thus, our algorithms consists of two phases; namely the self-play phase and the punishment phase. One challenge here is that the agents do not have any information about the payoff matrices of the stage game $\mathcal{G}$. Thus, the self-play phase itself needs to consist of two sub-phases called the exploration sub-phase and the exploitation sub-phase, each of which is described below. 

\textbf{Exploration Sub-Phase:} Since the payoff matrices are not completely known at the start, our algorithms begin in the exploration sub-phase. In the exploration sub-phase, every agent maintains and updates a local estimate of the payoff matrix of the other agent. 
The exploration sub-phase ends once the payoff matrices of both the agents are completely known to both the agents.
Since the exploration sub-phase is different for both our algorithms, we defer the specific details to later when we describe the algorithms. 

\textbf{Exploitation Sub-Phase:} The agents enter this sub-phase once the exploration sub-phase ends and if no agent has deviated from the algorithm during the exploration sub-phase. In this sub-phase, our first (resp. second) algorithm selects action according to fictitious play (resp. regret-matching) and remains in this phase until it detects that the other agent has deviated from fictitious play (resp. regret-matching). 

\textbf{Punishment phase:} An agent enters this phase if it detects that the other agent has 
deviated from the prescribed learning algorithm. Note that this means that the agent may enter the punishment phase either from the exploration sub-phase or from the exploitation sub-phase. Since the punishment strategy is common to both our algorithms, we now describe the punishment strategy.

We start with the definition of the \emph{minimax} strategy which is used in the punishment phase.

\begin{definition}\label{def:minimax}
The minimax value for agent $1$ on some matrix $Q$ is defined as
\begin{equation}\label{eq:minimax}
    \Bar{V}_1(Q) = \min_{y\in\mathcal{Y}} \max_{z\in\mathcal{Z}} y^{\top}Qz,
\end{equation}
where 
\begin{align*}
    \mathcal{Y}:= \{y\in \mathbb{R}^{|A_1|}:\sum_j y_j =1, \quad y_j\geq 0\forall j \},\\
    \mathcal{Z}:= \{z\in \mathbb{R}^{|A_2|}:\sum_k z_k =1, \quad z_k\geq 0\forall k \},
\end{align*}    
and the corresponding policy $y^*$ is called the minimax strategy for agent $1$. The definitions for the minimax value and strategy for agent $2$ are analogous.
\end{definition}

In what follows, without loss of generality, we will assume that agent $1$ deviates and refer to it as the adversary. Further, we will denote the local estimate of the payoff matrix of agent $1$ that agent $2$ maintains as $\hat{R}^1$. Note that, when the payoff matrix of agent $1$ is completely known by agent $2$, $\hat{R}^1=R^1$.
In the punishment phase, the idea is to \emph{punish} the adversary for not adhering to the algorithm. Since the adversary can deviate from the algorithm either during the exploration sub-phase or the exploitation sub-phase, the punishment strategy depends on when the adversary deviates. We will first describe the punishment strategy when an adversary deviates during the exploitation sub-phase followed by when it deviates during the exploration sub-phase. We summarize the punishment strategy in Algorithm \ref{alg:punish}.

Let $t$ denote the time when the punishment phase begins. Recall that in the exploitation sub-phase, the payoff matrices $\hat{R}^1$ and $R^2$ are completely known by agent $2$. Therefore, if the adversary deviates during the exploitation sub-phase, the punishment strategy is to select an action for agent $2$ by computing the minimax strategy, defined in Definition \ref{def:minimax}, on matrix $\hat{R}^1$ and execute it for all time $\tau\geq t$. On the other hand, if the adversary deviates during the exploration sub-phase, since the payoff matrix $R^1$ of the adversary is not completely known to agent $2$, the minimax strategy on $\hat{R}^1$ cannot be computed. Therefore, the idea is to replace the payoff matrix $\hat{R}^1$ with a different payoff matrix $\tilde{R}^1$ and execute a minimax strategy on $\tilde{R}^1$. Specifically, agent $2$ constructs a payoff matrix $\tilde{R}^1$ corresponding to the local estimate of the adversary's original payoff matrix $\hat{R}^1$ such that the entries that are known in $\hat{R}^1$ are the same in $\tilde{R}^1$. For the entries that are not known in $\hat{R}^1$, agent $2$ substitutes those entries as $0$ in $\tilde{R}^1$. Then, agent $2$ selects an action with equal probability of $\frac{1}{|A_2|}$ until at least one of the rows of matrix $\hat{R}^1$ is completely known, updating the unknown entries matrix $\tilde{R}^1$ as they are revealed to the agent. Note that in case agent $2$ deviates instead of agent $1$, then agent $1$ selects an action with equal probability of $\frac{1}{|A_1|}$ until at least one of the columns of matrix $\hat{R}^2$ is completely known.
Once at least one of the rows of matrix $\hat{R}^1$ is completely known,  agent $2$ then computes and executes the minimax strategy, defined in Defintion \ref{def:minimax}, on matrix $\tilde{R}^1$. If no new entry of matrix $R^1$ is revealed, the agent continues to play the computed minimax strategy. Otherwise, agent $2$ updates $\hat{R}^1$ and $\tilde{R}^1$ and recomputes the minimax strategy on $\tilde{R}^1$. 


\begin{lemma}\label{lem:punishment}
    Let $\mathcal{A}$ denote an algorithm that consists of the punishment strategy defined in Algorithm \ref{alg:punish} and without loss of generality, suppose agent $1$ deviates from $\mathcal{A}$. Further, let $\bar{V}_1^p:=\min_{a_{2}\in A_{2}}\max_{a_1\in A_1} r^1_{a_1,a_2}$.
    Then, algorithm $\mathcal{A}$ is perfectly rational if the following condition holds:
    \begin{equation}\label{eq:minimax_pure}
        \bar{V}_1^p\leq U_1(\mathcal{A},\mathcal{A}).
    \end{equation}
\end{lemma}

Lemma \ref{lem:punishment} ensures that by using the punishment strategy described in this section, we can design learning algorithms that are perfectly rational. However, we have not described how a learning algorithm detects if an agent has deviated or not.
We now present two algorithms that can detect any deviation by an agent and incorporate the punishment strategy described in Algorithm \ref{alg:punish}. Note that, for ease of understanding, we will describe the algorithms for agent $2$. The description for agent $1$ is analogous. 

\begin{algorithm}[tb]
   \caption{Rational Fictitious Play (R-GFP)}
   \label{alg:Sec-FP}
\begin{algorithmic}
   \STATE Select first action in $A_i$.
   \FOR{time $t>1$}
   \IF{in exploration sub-phase}
   \STATE Compute $E_i^t$ and select $a_i^t = BR_i^{\text{Fict}}(\hat{\textbf{a}}_{-i}(t))$ on $E_t^i$.
   \ELSE
   \STATE Select $a_i^t = BR_i^{\text{Fict}}(\hat{\textbf{a}}_{-i}(t))$ on $R^i$.
   \ENDIF
   \STATE Observe the payoffs and the actions.
   \IF{Agent $-i$ deviated}
   \STATE Enter Punishment Phase (Algorithm \ref{alg:punish}).
   \ENDIF
   \ENDFOR
\end{algorithmic}
\end{algorithm}
\subsection{Rational Generalized Fictitious Play}
In this section, we describe our first algorithm Rational Generalized Fictitious Play (R-GFP). We start with the following definition that generalizes the fictitious play algorithm described in Section \ref{sec:model}.

\begin{definition}[Generalized Fictitious Play (GFP)]\label{def:GFP}
    For a repeated game formed through iterations of the stage game $\mathcal{G}$, let $H_2(t)$ denote the sequence of actions selected by agent $2$ until time $t$. Let $h_2(t)\subseteq H_2(t)$ denote any subset of $H_2(t)$. Then, the best-response $BR(\cdot)$ for agent $1$ at time $t+1$ is defined as
    \begin{equation}\label{eq:GFP}
        BR_1^{\text{GFP}}(h_2(t))\coloneqq \arg\max_{a\in A_1}R^1(a,\hat{\textbf{a}}_2(h_2(t))),
    \end{equation}
    where $\hat{\textbf{a}}_2(h_2(t))$ denotes the vector of empirical frequencies of actions $a_2\in A_{2}$ determined by using $h_2(t)$.
\end{definition}
The definition for agent $2$ is analogous. Note that when $h_2(t)=H_2(t)$, we retrieve the fictitious play algorithm. Algorithm R-GFP is summarized in Algorithm \ref{alg:Sec-FP} and is described as follows.

\begin{figure}[t]
\centering
  \begin{subfigure}[b]{0.45\columnwidth}
  \centering
    \includegraphics[width=\linewidth]{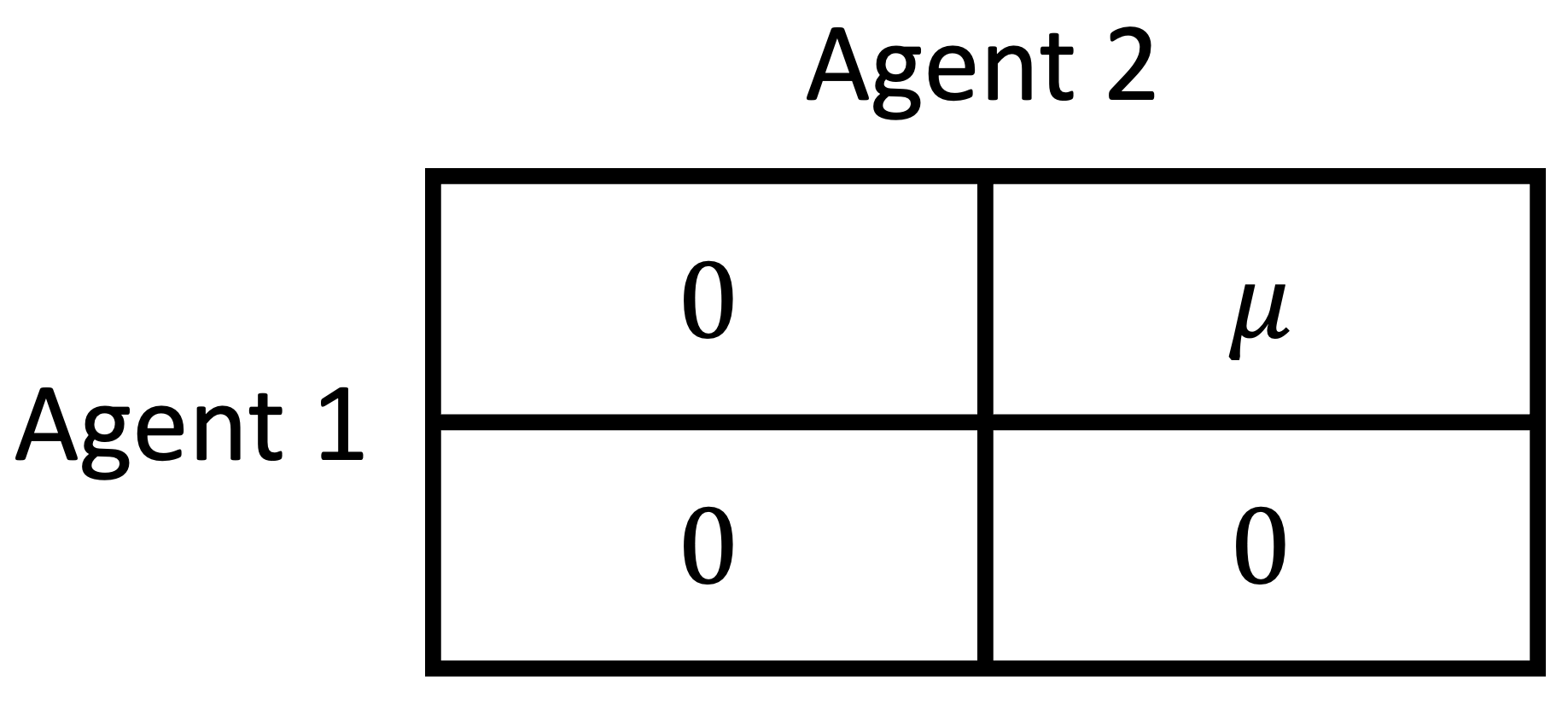}
    \caption{Time $t=2$.}
  \end{subfigure}
  \hfill 
  \begin{subfigure}[b]{0.45\columnwidth}
  \centering
    \includegraphics[width=\linewidth]{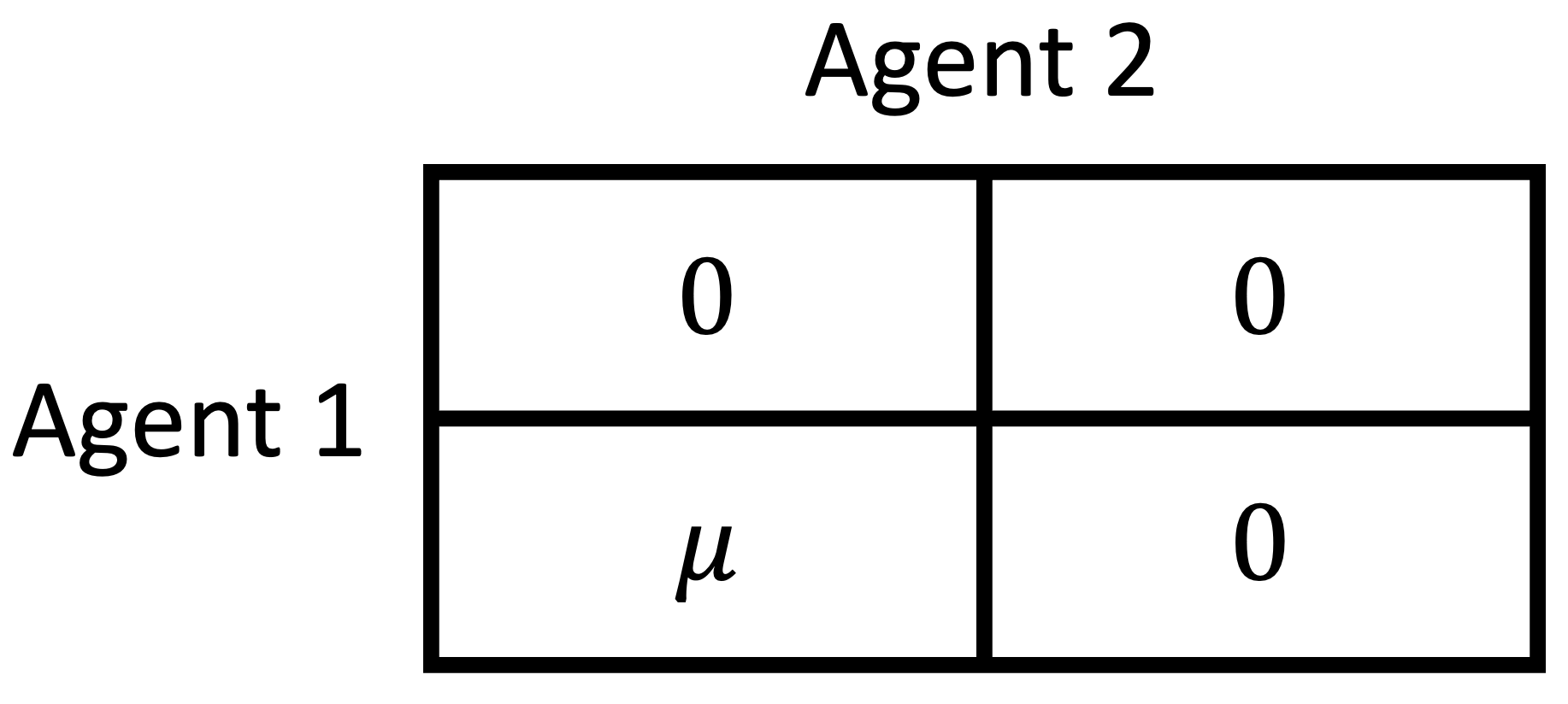}
    \caption{Time $t=3$.}
  \end{subfigure}
  \caption{Construction of matrix $E_2^t$ in Algorithm R-GFP at times $t=2$ and $t=3$.}
  \label{fig:Detect_FP}
\end{figure}

The exploration sub-phase begins at time $t=1$. At the first time instant, Algorithm R-GFP selects the first action in the set $A_2$ for agent $2$. If any entry except the $(1,1)$ entry of matrix $R^2$ is revealed, then that means that agent $1$ has deviated from the algorithm and the algorithm enters the punishment phase. Otherwise, Algorithm R-GFP continues as follows. At each time $1< t \leq |A_1||A_2|$, agent $2$ constructs a matrix $E_2^t$ with all of the entries as zero except the $(j,k)$ entry, where $j=q$, $k=t-(q-1)|A_2|$, $q:= \lceil\frac{t}{|A_2|}\rceil$, and $\lceil\cdot\rceil$ denotes the ceil function. The $(j,k)$ entry of the payoff matrix $E_2^t$ is set to $\mu$, where $\mu$ is any positive real number\footnote{The quantity $\mu$ need not be same for both agents.}. Then, agent $2$ selects an action according to fictitious play, defined in equation \eqref{eq:FP}, on $E_2^t$. Agent $2$ then observes its own and agent $1$'s payoff and updates $\hat{R}^1$ and $R^2$. If at time $t$, any entry other than $(j,k)$ entry of $R^2$ is revealed, Algorithm R-GFP enters the punishment phase.
Figure \ref{fig:Detect_FP} provides an illustration of matrix $E_2^t$ for different values of time $t$.


Once the exploration sub-phase ends, i.e., at time $t=|A_1||A_2|+1$, Algorithm R-GFP enters the exploitation sub-phase in which agent $2$ selects its actions according to the generalized fictitious play defined in equation \ref{eq:GFP}.  Agent $2$ then observes the payoffs obtained by both agents. Given that the payoffs matrices are completely known in this sub-phase, agent $2$ then determines whether agent $1$ deviated or not. Mathematically, agent $2$ checks whether the following condition holds at time $t$.
\begin{equation}\label{eq:FP_check}
    a_{1}^t=BR_{1}^{\text{GFP}}(h_2(t)).
\end{equation}
If equation \eqref{eq:FP_check} holds at time $t$, then agent $2$ continues to select action according to equation \eqref{eq:FP} for time $t+1$. Otherwise, Algorithm R-GFP enters the punishment phase.

To determine if agent $1$ deviated at time $t$, agent $2$ requires the information of $h_2(t)$ that agent $1$ considers. This can be achieved by keeping $h_i(t), i\in\{1,2\}$ for both agents same, such as in fictitious play.


\begin{theorem}\label{thm:Sec_FM}
Let $\pi_1^*$ and $\pi_2^*$ denote the strategies of agent $1$ and agent $2$, respectively, if they selected actions according to generalized fictitious play defined in Definition \ref{def:GFP}. Then, 
    \begin{enumerate}
        \item Algorithm R-GFP is perfectly rational if the condition defined in equation \eqref{eq:minimax_pure} holds.
        \item If $\pi_1^*$ and $\pi_2^*$ converge to an equilibrium then, the strategies of the agents obtained when they follow Algorithm R-GFP also converge to an equilibrium.
    \end{enumerate} 
\end{theorem}

\begin{algorithm}[tb]
   \caption{Rational Regret Matching (R-RM)}
   \label{alg:Sec-RM}
\begin{algorithmic}
    \STATE {\bfseries Input:} $\delta\in(0,1)$.    
   \STATE Select first action in $A_i$.
   \FOR{epoch $t>1$}
   \IF{in exploration sub-phase}
   \STATE Compute $E_t^i$ and select $a_i^t$ according to equation \eqref{eq:sec_rm_prob}.
   \ELSE
   \STATE Compute the probability distribution over actions $a_i$ according to equation \eqref{eq:rm_prob}.
   \STATE Compute $N_t$ and $\epsilon_t$.
   \FOR{$n\leq N_t$}
   \STATE Select $a_i^t$ according to the computed probability distribution.
   \STATE Observe agent $\mathcal{P}_{-i}$'s actions.
   \ENDFOR
   \ENDIF
   \IF{Agent $\mathcal{P}_{-i}$ deviated}
   \STATE Enter Punishment Phase (Algorithm \ref{alg:punish}).
   \ENDIF   
   \ENDFOR
\end{algorithmic}
\end{algorithm}

\begin{algorithm}[tb]
   \caption{Strategy for Punishment Phase}
   \label{alg:punish}
\begin{algorithmic}
    \STATE Assumes agent $1$ deviates.
    \FOR{time $t\geq 1$}
    \IF{matrix $\hat{R}^1$ is completely known to agent $2$}
    \STATE Select minimax strategy (Defintion \ref{def:minimax}) on $\hat{R}^1$.
    \ELSE
    \STATE Construct matrix $\tilde{R}^1$.
   \STATE Select minimax strategy (Defintion \ref{def:minimax}) on $\tilde{R}^1$.
   \STATE Update matrix $\hat{R}^1$ and $\tilde{R}^1$ if new entry is revealed. 
   \ENDIF
   \ENDFOR
\end{algorithmic}
\end{algorithm}

\subsection{Rational Regret Matching}
We now describe our second algorithm which, with high probability, is also perfectly rational. The idea behind Algorithm Rational Regret Matching (R-RM) is similar to that of Algorithm R-GFP. Specifically, during the exploitation sub-phase, agent $2$ selects an action according to regret-matching described in Section \ref{sec:model} and checks whether agent $1$ has deviated or not. However, recall from Section \ref{sec:model}, that there is a probability distribution over the actions in regret matching algorithm. Thus, to determine whether agent $1$ has  deviated, agent $2$ must compare the probability distribution over the actions selected by agent $1$ with the probability distribution over the actions that agent $1$ must have selected its actions from. To achieve this, Algorithm R-RM runs in epochs consisting of a finite number of iterations. We summarize Algorithm R-RM in Algorithm \ref{alg:Sec-RM} and describe it in greater detail below.

In the first epoch, Algorithm R-RM selects the first action from the set $A_2$. If any entry except the $(1,1)$ entry of matrix $R^2$ is revealed, then the algorithm enters the punishment phase. Otherwise, the algorithm continues as follows.

In every epoch $t$, $1<t\leq |A_1||A_2|$, agent $2$ constructs a matrix $E_2^t$ in which, excluding at most two entries, all other entries are zero. Denote the two non-zero entries as $(j_1,k_1)$ and $(j_2,k_2)$. Let $q:= \lceil\frac{t}{|A_2|}\rceil$. Then, entry $(j_1,k_1)$ is set to $\mu>0$ and is defined as follows.
\begin{align}
    &j_1 = 
    \begin{cases}
        \max\{1,(q-1)\}, \text{ if } t=(q-1)|A_2|+1,\\
        q, \text{ otherwise.}
    \end{cases}\\
    &k_1 = \begin{cases}
        1, \text{ if } t=(q-1)|A_2|+1,\\
        t-|A_2|(q-1), \text{ otherwise.}
    \end{cases}
\end{align}
Entry $(j_2,k_2)$ is set to $\nu>0$ if $t=(q-1)|A_2|+1$ and zero otherwise, where $j_2 = q$ and $k_2 = |A_2|$.
\begin{figure}[t]
\centering
  \begin{subfigure}[b]{0.45\columnwidth}
  \centering
    \includegraphics[width=\linewidth]{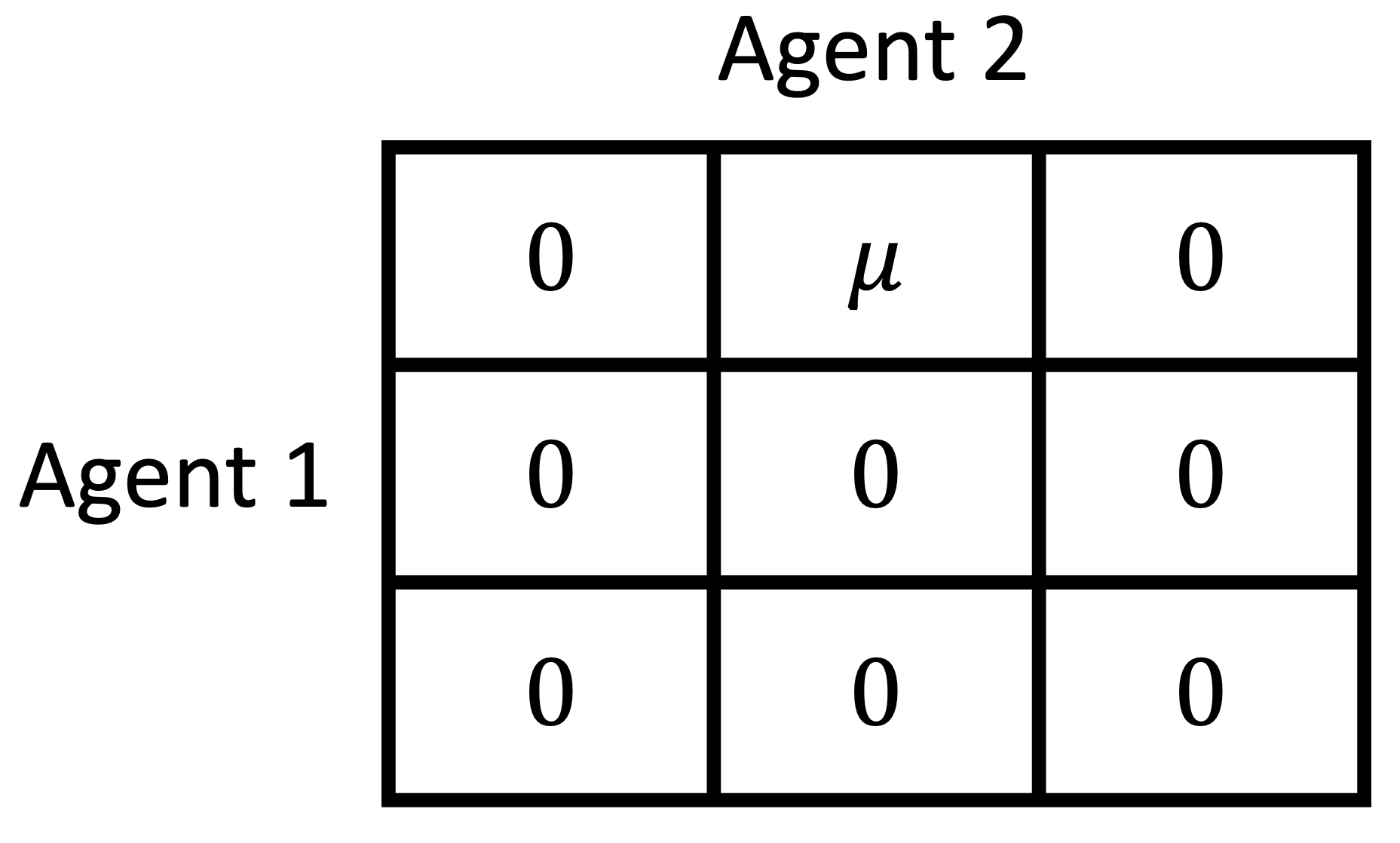}
    \caption{Time $t=2$.}
  \end{subfigure}
  \hfill 
  \begin{subfigure}[b]{0.45\columnwidth}
  \centering
    \includegraphics[width=\linewidth]{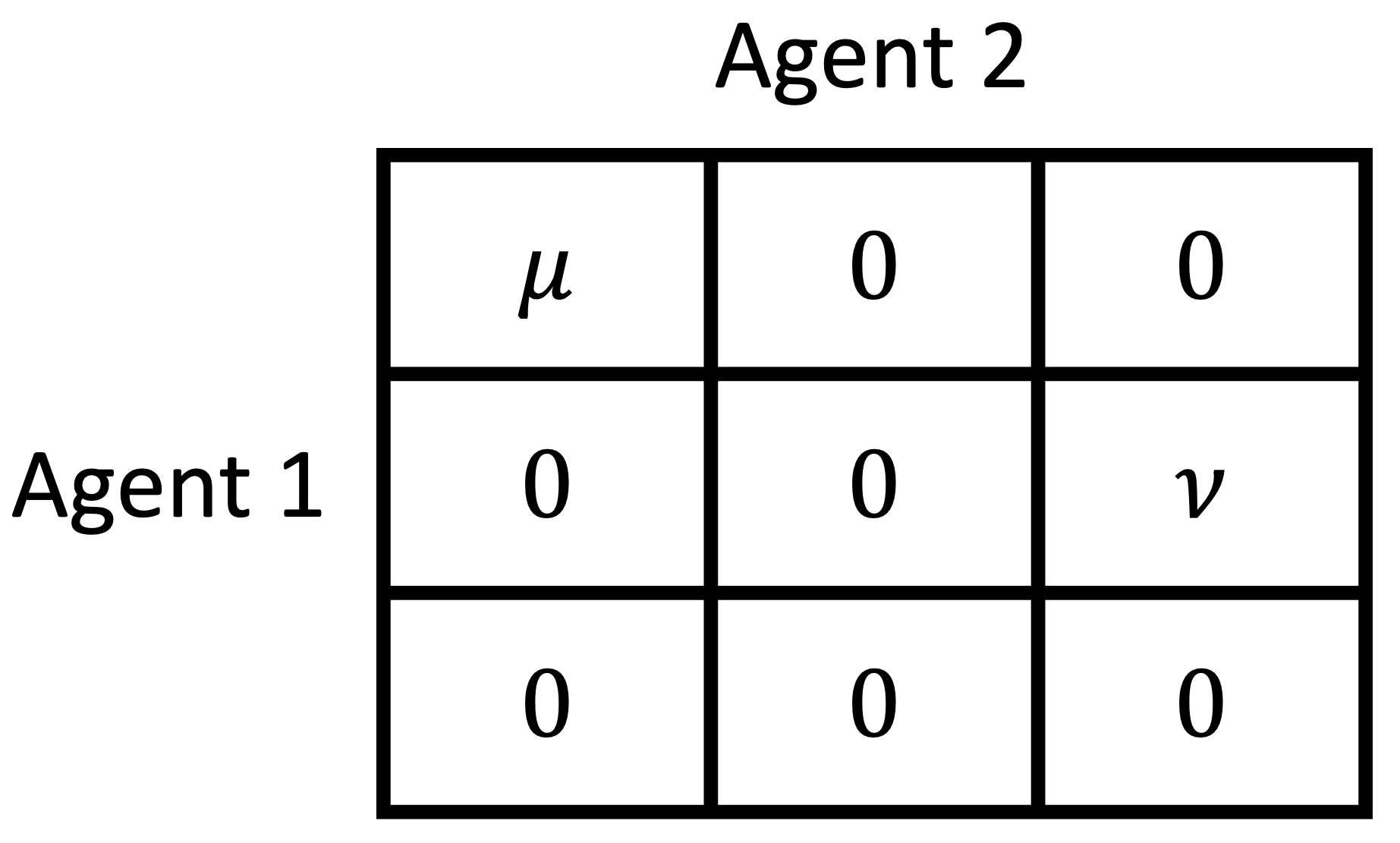}
    \caption{Time $t=4$.}
  \end{subfigure}
  \caption{Construction of $E_2^t$ at epoch $t=2$ and $t=4$ for Algorithm R-RM.}
  \label{fig:detect_RM}
\end{figure}
Figure \ref{fig:detect_RM} provides an illustration of matrix $E_2^t$ for different epochs $t$. Once matrix $E_2^t$ is determined, agent $2$ selects an action proportional to the instantaneous regret determined on $E_2^t$. Formally, the instantaneous regret for epoch $t$ for agent $2$ for action $a\in A_2$ is $\delta_2^t(a) = e^2_{a_1^t,a}-e^2_{a_1^t,a_2^t}$, where $e^{2}_{i,j}$ denotes the $(i,j)$ entry of $E_2^t$.
Agent $2$ then selects action $a\in A_2$ with probabilities
\begin{align}\label{eq:sec_rm_prob}
    p_{t+1}^2(a) = 
    \begin{cases}
        \frac{\delta_+(a)}{\sum_{a'\in A_2}\delta_+(a')}, \text{ if } \sum_{a'\in A_2}\delta_+(a')>0,\\
        a_2(t-1), \text{ otherwise,}
    \end{cases}
\end{align}
where $\delta_+(a) \coloneqq \max\{0,\delta_{2}^t(a)\}$.
Similar to Algorithm R-GFP, the exploration sub-phase in Algorithm R-RM ensures that the payoffs of the original matrices is revealed sequentially to the agent. Thus, if agent $1$ deviates from Algorithm R-RM during the exploration sub-phase, it is guaranteed to be detected by agent $2$. Once the exploration phase ends, the Algorithm R-RM enters the exploitation sub-phase. 

During the exploitation sub-phase, at the start of  epoch $t> |A_1||A_2|+1$ and since the payoff matrices are completely known, Algorithm R-RM determines the probability distribution over the actions of agent $2$ as well as agent $1$ by using equation \eqref{eq:rm_prob}. Let $\phi_{t}^1$ and $\phi_{t}^2$ denote the probability distribution over the actions from which agent $1$ and agent $2$, respectively, must choose their actions from, if they were following the regret matching algorithm.
Every epoch $t$ consists of $N_t$ iterations, where $N_t$ will be determined shortly. In every iteration $n\leq N_t$ of epoch $t$, agent $2$ selects action according to $\phi^2_t$, and observes the action selected by agent $1$. Once the epoch ends, i.e., after $N_t$ iterations, agent $2$ determines the empirical cumulative distribution function (CDF) from the observed actions of agent $1$. Formally, let $a_l, l\in\{1,\dots,N_t\}$, be the actions of agent $1$ observed by agent $2$ in epoch $t$. Then, the empirical CDF $F_{t}^1$ over the actions of agent $1$ computed by agent $2$ after epoch $t$ is defined as
\begin{equation}\label{eq:empirical_CDF}
    F_t^1(x) = \frac{1}{N_t}\sum_{i=1}^{N_t}\textbf{1}_{\{a_i\leq x\}},~ x\in\mathbb{R}.
\end{equation}
Let $\mathcal{F}_t^1(x)$ denote the CDF determined using $\phi_{t}^1(a)$. Then, after computing the empirical CDF, Algorithm R-RM checks whether the following condition holds at the end of epoch $t$:
\begin{align}\label{eq:CDF_comp}
    \sup_{x\in \mathbb{R}} |\mathcal{F}_t^1(x)-F_t^1(x)| > \epsilon_t,
\end{align}
where $\epsilon_t=\tfrac{1}{t}$.
If condition \eqref{eq:CDF_comp} holds, Algorithm R-RM proceeds to the next epoch $t+1$. Otherwise, Algorithm R-RM enters the punishment phase. 

Algorithm R-RM sets $N_t = \frac{c_1\log(\frac{c_2t}{\delta})}{\epsilon^2_t}$, where $\delta\in(0,1)$ is an input to Algorithm R-RM and $c_1>0$ and $c_2>1$ are some real numbers satisfying $2\geq c_2t^{2c_1-1}$. Although the choice of $N_t$ will be clear from the proof of Theorem \ref{thm:Sec_RM}, we provide an intuition behind this choice. On one hand, we require that in case agent $1$ does not deviate from Algorithm R-RM, the equilibrium strategies of Algorithm R-RM converge to that of when the agents would have selected actions according to the regret matching algorithm. To achieve this, we must ensure that the condition defined in equation \eqref{eq:CDF_comp} holds with very low probability (almost $0$), when agent $1$ does not deviate. On the other hand, in case agent $1$ deviates, we require that Algorithm R-RM enters the punishment phase. Thus, motivated from \cite{conitzer2007awesome}, we select $N_t$ (resp. $\epsilon_t$) such that it increases (resp. decreases) in every epoch $t$. 


\begin{theorem}\label{thm:Sec_RM}
    Let $\pi_1^*$ and $\pi_2^*$ denote the strategies of agent $1$ and agent $2$, respectively, if they selected actions according to regret-matching. Then, for a given $\delta\in (0,1)$,
    \begin{enumerate}
        \item Algorithm R-RM is perfectly rational if the condition defined in equation \eqref{eq:minimax_pure} holds..
        \item If $\pi_1^*$ and $\pi_2^*$ converge to an equilibrium then, with probability $1-\delta$, the strategies of the agents converge to an equilibrium if they follow Algorithm R-RM.
    \end{enumerate} 
\end{theorem}

We now conclude this work by establishing that, for any given constant $c$, a $c$-rational algorithm may not always exist in case of \emph{imperfect monitoring}, i.e., when an agent can observe the actions of the other agent but not its payoffs. 
\begin{theorem}\label{thm:imperfect_games}
    Under imperfect monitoring, there exist classes of games for which no algorithm is $c$-rational for any $c\geq 1$.
\end{theorem}

For certain classes of games such as zero-sum games or common interest games, a $c$-rational algorithm may exist under imperfect monitoring. This is because, in these classes of games, the payoff of the agent $i$ provides information about the payoffs of agent $-i$ as well. 

\section{Numerical Results}\label{sec:numerics}
We now provide numerical results to illustrate the analytical results established in Theorems \ref{thm:Sec_FM} and \ref{thm:Sec_RM}. Note that when agent $i$ deviates during the exploration (resp. exploitation) sub-phase, then the algorithm followed by agent $i$ is denoted as $\mathcal{A}'_{\text{explore}}$ (resp. $\mathcal{A}'$). The constants $c_1$, $c_2$, and $\delta$ were selected to be $\tfrac{9}{8}$, $8$, and $0.01$, respectively. Further, all of our numerical results represents the mean over $50$ runs.

\begin{figure}[t]
    \centering
    \includegraphics[scale=0.18]{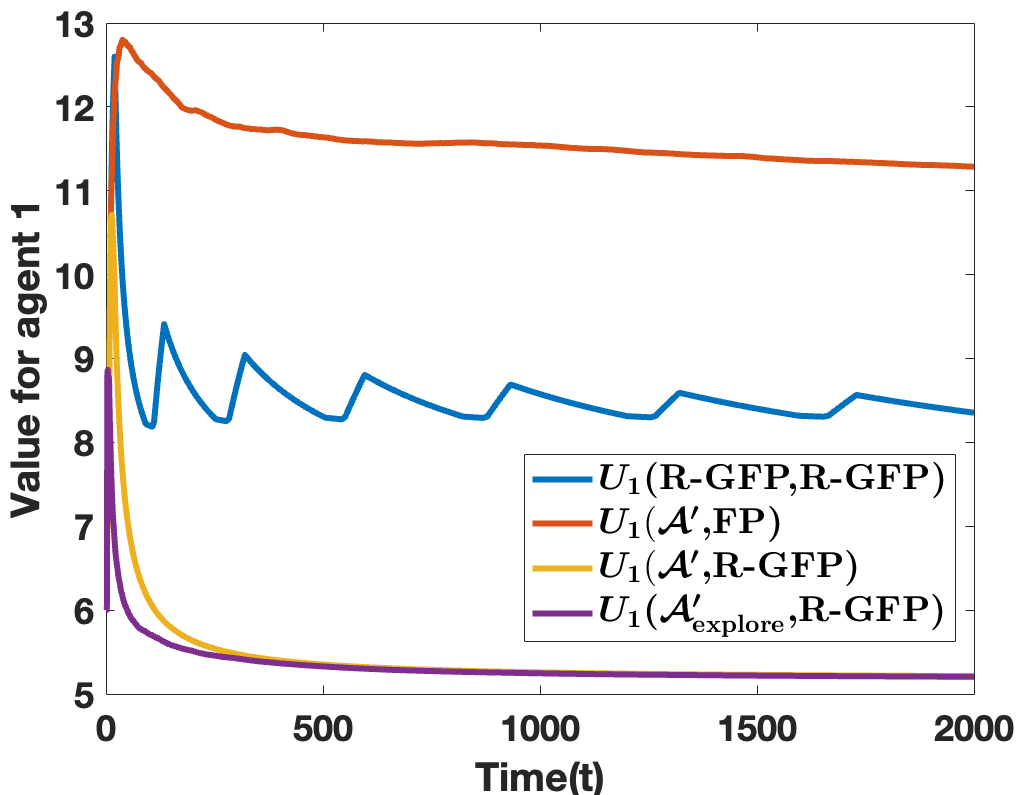}
    \caption{Numerical plot illustrating value of agent $1$ (adversary) over time.}
    \label{fig:RFP_plot}
\end{figure}
\begin{figure}[t]
    \centering
    \includegraphics[scale=0.33]{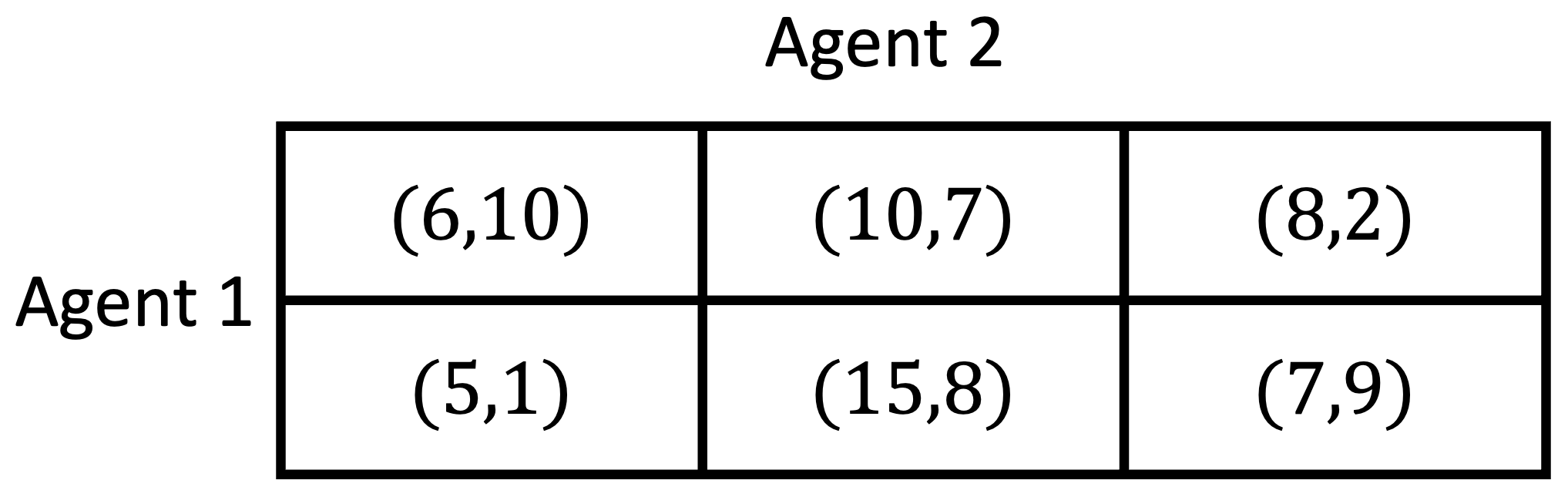}
    \caption{Game used for the numerical results in Figure \ref{fig:RFP_plot}.}
    \label{fig:RFP_plot_game}
\end{figure}

Figure \ref{fig:RFP_plot} illustrates the value of agent $1$ in four cases: (a) when both agents follow Algorithm R-GFP, (b) when agent $1$ deviates and agent $2$ follows fictitious play (FP), (c) when agent $1$ deviates during exploitation sub-phase and agent $2$ follows Algorithm R-GFP, and (d) when agent $1$ deviates during exploration sub-phase and agent $2$ follows Algorithm R-GFP. In all of these cases, the agents utilize the entire history, i.e., $h_i(t)=H_i(t)$, and thus, the generalized fictitious play is equivalent to fictitious play (FP). An optimal strategy for an adversary against fictitious play was determined in \cite{vundurthy2023intelligent}. Thus, upon deviation, agent $1$ follows the strategy described in \cite{vundurthy2023intelligent}. The game is illustrated in Figure \ref{fig:RFP_plot_game} and has three Nash equilibria with Nash outcomes $(6,10)$, $(8.33,7.83)$, and $(6.66,7.3)$.

The strategies obtained for agent $1$ and agent $2$ when they both follow Algorithm R-GFP is $\begin{bmatrix} 0.167 & 0.833 \end{bmatrix}^{\top}$ and $\begin{bmatrix} 0 & 0.167 & 0.833 \end{bmatrix}^{\top}$, respectively, corresponding to the Nash outcome of $(8.3,7.83)$. The curve obtained for $U_2(\text{FP},\text{FP})$ is analogous to that of $U_2(\text{R-GFP},\text{R-GFP})$ and thus has not been shown in Figure \ref{fig:RFP_plot}.
From Figure \ref{fig:RFP_plot}, the value $U_1(\mathcal{A}',\text{FP})$ of agent $1$ is higher than that of $U_1(\text{FP},\text{FP})$ meaning that there is benefit for agent $1$ to deviate from fictitious play. However, $U_1(\mathcal{A}',\text{R-GFP})<U_1(\text{R-GFP},\text{R-GFP})$ meaning that there is no benefit for agent $1$ to deviate from Algorithm R-GFP as characterized in Theorem \ref{thm:Sec_FM}.
Finally, note that for low values of time, $U_1(\mathcal{A}'_{\text{explore}},\text{R-GFP})<U_1(\mathcal{A}',\text{R-GFP})$ and as $t\to\infty$, $U_1(\mathcal{A}'_{\text{explore}},\text{R-GFP})\to U_1(\mathcal{A}',\text{R-GFP})$. This is because, as time increases, matrix $R^1$ is completely known by agent $1$ and thus, the minimax value is same in both cases.

\begin{figure}[t]
    \centering
    \includegraphics[scale=0.18]{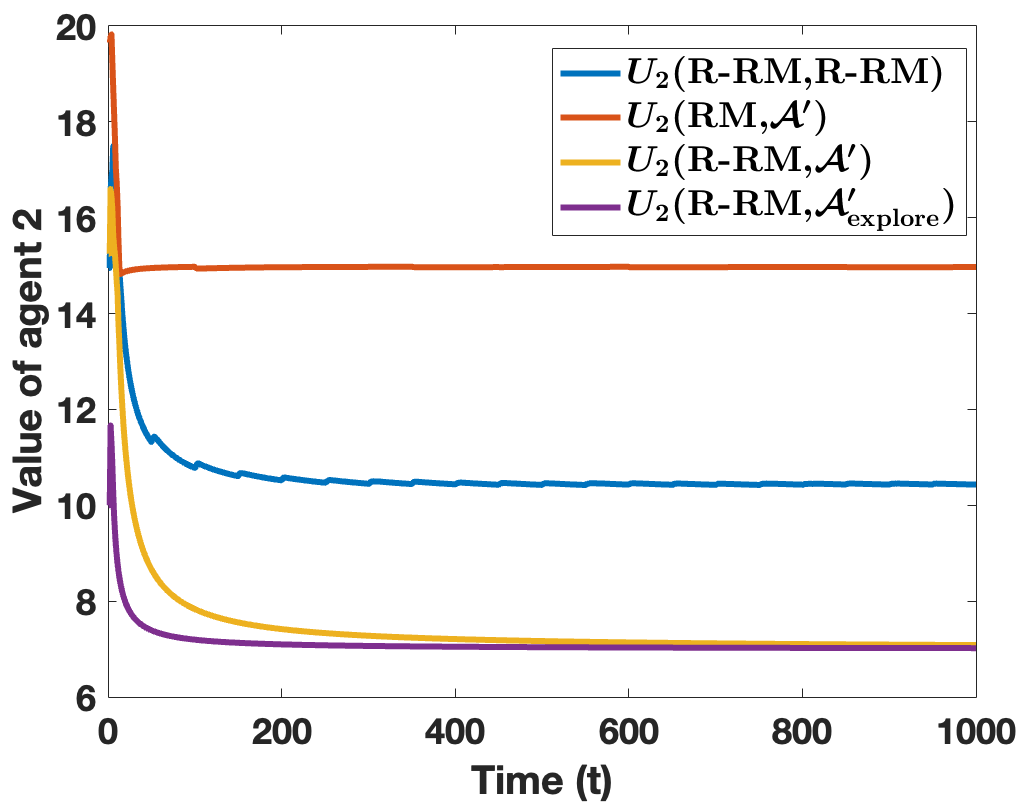}
    \caption{Numerical plot illustrating value of agent $2$ (adversary) over time. regret-matching is denoted as RM.}
    \label{fig:RRM_plot}
\end{figure}
\begin{figure}[t]
    \centering
    \includegraphics[scale=0.33]{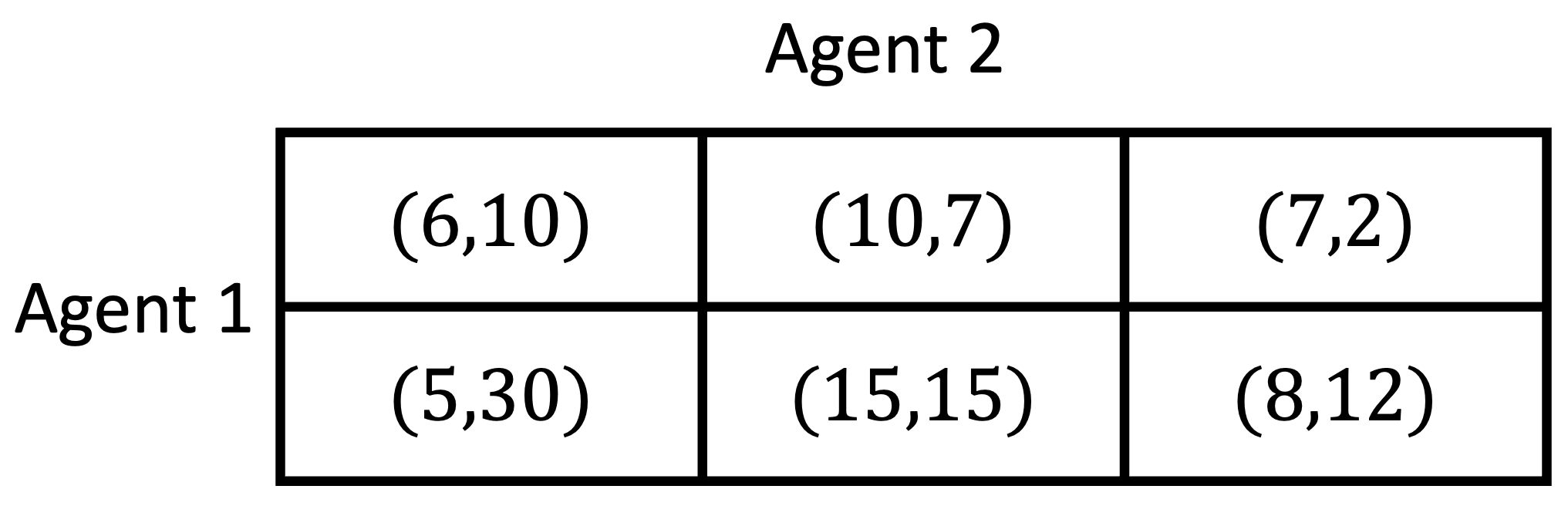}
    \caption{Game used for numerical result in Figure \ref{fig:RRM_plot}.}
    \label{fig:RRM_plot_game}
\end{figure}

Figure \ref{fig:RRM_plot} illustrates the value of agent $2$ in four cases as described for Figure \ref{fig:RFP_plot}. Upon deviation, agent $2$ follows the strategy described as follows: 
\begin{equation}\label{eq:strat_adv_RM}
    a_2^* = \argmax_{a_2\in A_2} R^2(BR(a_2),a_2),
\end{equation}
where $BR(a_2):= \argmax_{a_1\in A_1} R^1(a_1,a_2)$.
Note that the strategy defined in equation \eqref{eq:strat_adv_RM} may not be optimal. For instance, $U_2(RM,\mathcal{A}')=14.9$, where $\mathcal{A}'$ is defined in equation \eqref{eq:strat_adv_RM}. However, by considering a mixed strategy as $\begin{bmatrix}
0.15 & 0.85 & 0\end{bmatrix}^{\top}$ for agent $2$ yields a higher payoff of $17.3$. When agent $1$ follows regret-matching (RM) and agent $2$ deviates, $U_2(\text{RM},\mathcal{A}')>U_2(\text{RM},\text{RM})$ implying that by deviating from regret-matching, agent $2$ can obtain a higher value. Since $U_2(\text{RM},\text{RM})=U_2(\text{R-RM},\text{R-RM})$, the curve for $U_2(\text{RM},\text{RM})$ is not shown in Figure \ref{fig:RRM_plot}. However, $U_2(\text{R-RM},\mathcal{A}')<U_2(\text{R-RM},\text{R-RM})$ and $U_2(\text{R-RM},\mathcal{A}'_{\text{explore}})<U_2(\text{R-RM},\text{R-RM})$ implying that Algorithm R-RM is perfectly rational. Finally, since the strategy followed by agent $2$ does not change once agent $1$ follows minimax strategy, $U_2(\text{R-RM},\mathcal{A}'_{\text{explore}})$ is approximately the same as $U_2(\text{R-RM},\mathcal{A}')$.

We provide additional numerical results in the Appendix for various games to illustrate the results for both Algorithm R-GFP and Algorithm R-RM.

\section{Conclusion}\label{sec:conclusion}
In this work, we considered a two-agent non-cooperative repeated game framework and introduced rationality ratio, which is a ratio of the most an agent can obtain by deviating from a learning algorithm to their payoff from following it under an assumption of perfect monitoring. A learning algorithm is called $c$-rational if its rationality ratio is at most $c$. We first established that fictitious play and regret-matching algorithm are not $c$-rational for any given constant $c$ and then presented two algorithms that are provably $1$-rational. Finally, we established that there exists classes of games in which a $c$-rational algorithm does not exist under imperfect monitoring. 

Our algorithms extends to the case of $n$ agents when at most one agent can deviate. If more than one agent can deviate, the possibility of collusion arises which will be considered in the future. Apart from designing $c$-rational algorithms under imperfect monitoring (for classes of games in which they might exist), generalization to stochastic games is another potential future direction of this work.

\section{Impact Statement}
This paper presents work whose goal is to advance the field of non-cooperative multi-agent reinforcement learning. There are many potential societal consequences of our work. For instance, given the importance of learning algorithms in digital and financial markets, this work is useful in reducing the exploitability of learning algorithms in strategic situations.

\bibliography{main}
\bibliographystyle{icml2024}

\newpage
\appendix
\onecolumn
\section{Proof of Theorem \ref{thm:BR_not_secure}}
\begin{figure}[h]
    \centering
    \includegraphics[scale=0.4]{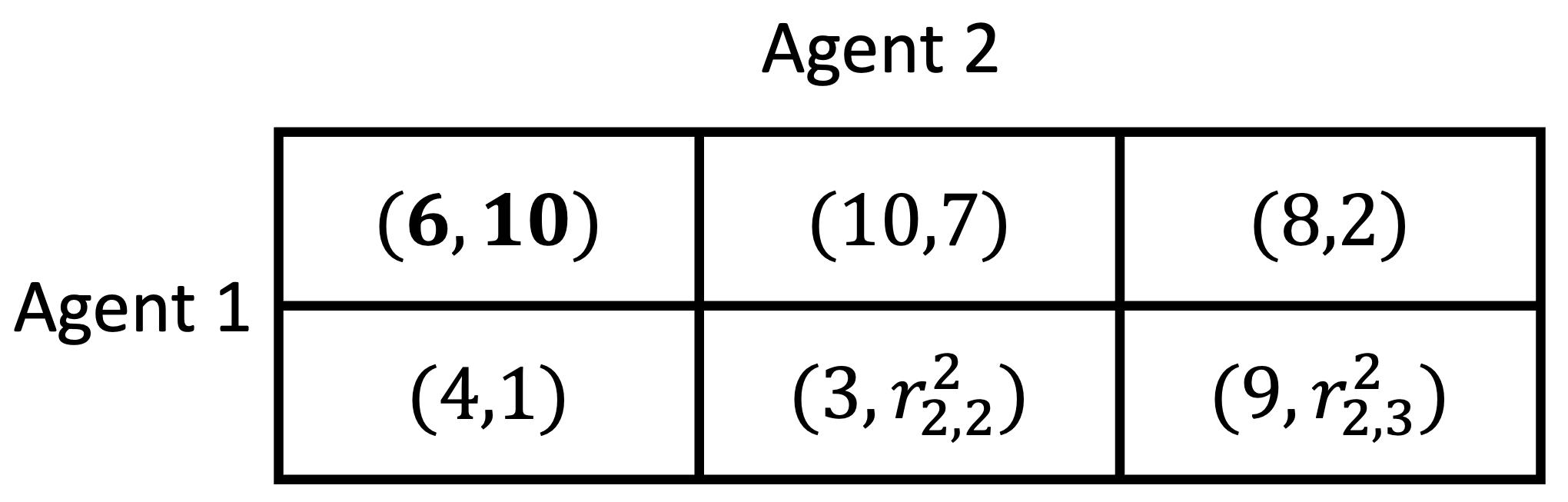}
    \caption{A $2\times3$ game $\mathcal{G}$ for the proof of Theorem \ref{thm:BR_not_secure}. Pure Nash equilibrium, highlighted in bold, is entry $(1,1)$.}
    \label{fig:BR_fargile}
\end{figure}
\begin{proof}
    Without loss of generality, suppose that agent $2$ deviates from fictitious play and let $\mathcal{A}'$ denote the algorithm followed by agent $1$. Further, let $\mathcal{A}$ denote the fictitious play algorithm.
    The idea behind this proof is to show that there exists a game $\mathcal{G}$ and algorithm $\mathcal{A}'$, such that the condition defined in equation \eqref{eq:secure_algo} does not hold for any given constant $c$. We start by describing a game $\mathcal{G}$.

    Consider a $2\times 3$ bi-matrix game $\mathcal{G}$, as illustrated in Figure \ref{fig:BR_fargile}. The payoff $r^2_{2,3}$ in $\mathcal{G}$ is set to $r^2_{2,3} = 10(c+1)$ and $r^2_{2,2}$ is any real number satisfying $r^2_{2,2}>r^2_{2,3}$. Note that entry $(1,1)$ corresponds to the  Nash equilibrium of $\mathcal{G}$. 
    Since fictitious play is known to converge to the Nash equilibrium in the class of non-degenerate ordinal potential games \cite{berger2005fictitious}, it is ensured that when both agents follow algorithm $\mathcal{A}$ on $\mathcal{G}$, the strategies converge to the Nash equilibrium. This implies that $U_2(\mathcal{A},\mathcal{A})=10$.
    
    We now describe algorithm $\mathcal{A}'$ that agent $2$ follows.
    Algorithm $\mathcal{A}'$, for each time $t$, has agent $2$ select column $k=3$. We will now show that, for algorithm $\mathcal{A}$ and a given constant $c$, equation \eqref{eq:secure_algo} does not hold.

    Let $H_2(t) = \{k,k,\dots,k\}$ denote the sequence of actions selected by agent $2$ until the current time $t$. As the action of agent $2$ does not change at any time $t$, it follows that $\hat{\textbf{a}}_2(t) = \begin{bmatrix} 0 & 0 & 1\end{bmatrix}$.
    Given game $\mathcal{G}$ as defined in Figure \ref{fig:BR_fargile} and since agent $1$ follows fictitious play, it follows that as $t\to\infty$, agent $1$ selects row $j=2$ as determined from equation \eqref{eq:FP}. This yields that $U_2(\mathcal{A},\mathcal{A}')=r^2_{2,3}$. Since $r^2_{2,3} = 10(c+1)$, it follows that
    \begin{align*}
        \frac{U_2(\mathcal{A},\mathcal{A}')}{U_2(\mathcal{A},\mathcal{A})} = c+1 \implies \sup_{\mathcal{G},\mathcal{A}'} \frac{U_2(\mathcal{A},\mathcal{A}')}{U_2(\mathcal{A},\mathcal{A})} \geq c+1.
    \end{align*}
    This further implies that the condition in equation \eqref{eq:secure_algo} can never hold for any given constant $c$. This concludes the proof.
\end{proof}
\section{Proof of Theorem \ref{thm:RM_not_secure}}
\begin{figure}[h]
    \centering
    \includegraphics[scale=0.4]{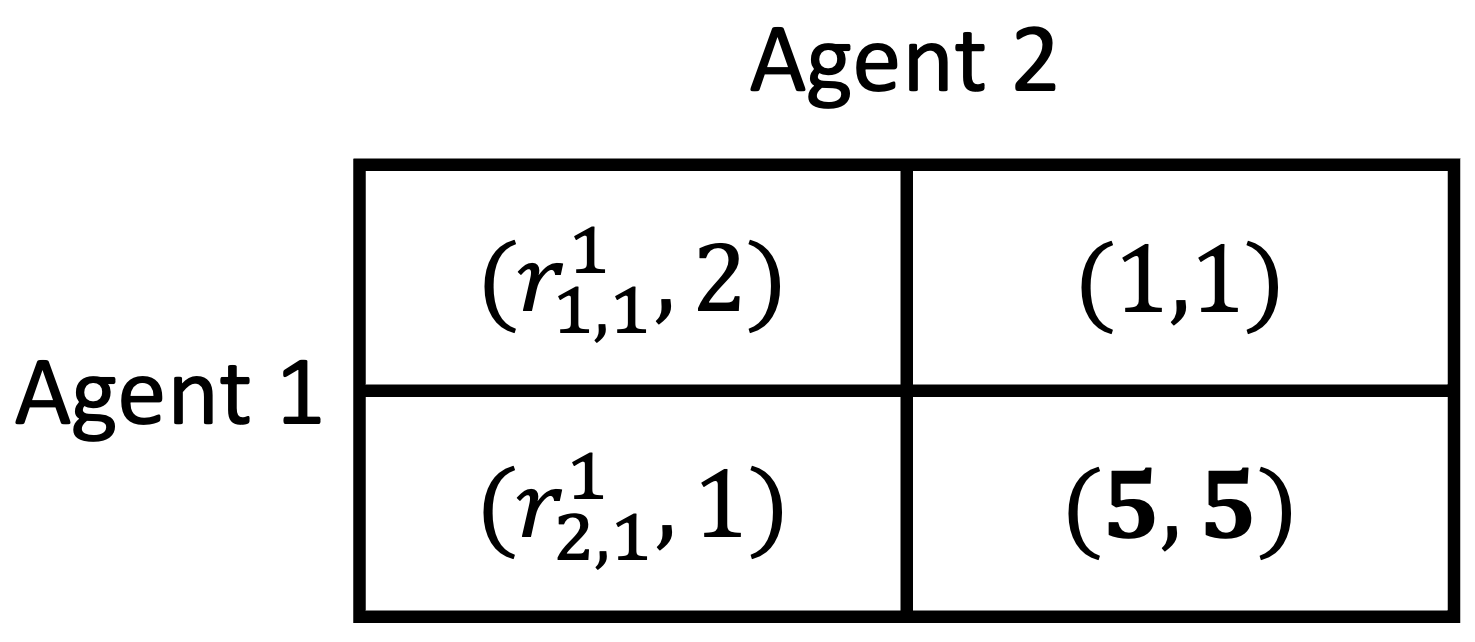}
    \caption{A $2\times2$ game $\mathcal{G}$ for the proof of Theorem \ref{thm:BR_not_secure}. For any $r^1_{2,1}>r^1_{1,1}$, entry $(2,2)$ is the Nash equilibrium.}
    \label{fig:RM_fragile}
\end{figure}
\begin{proof}
    Without loss of generality, suppose agent $1$ deviates from regret matching algorithm and let $\mathcal{A}'$ denote the algorithm followed by agent $1$. Further, let $\mathcal{A}$ denote the regret matching algorithm. Similar to the proof of Theorem \ref{thm:BR_not_secure}, we start by describing a game $\mathcal{G}$ followed by an Algorithm $\mathcal{A}'$ for agent $1$ on $\mathcal{G}$.

    Consider a $2\times 2$ bi-matrix game $\mathcal{G}$ as shown in Figure \ref{fig:RM_fragile} with entry $r^1_{1,1}=5(c+1)$ and $r^1_{2,1}>r^1_{1,1}$. It is known that no regret algorithms, such as regret matching, converge to the pure Nash equilibrium in general sum $2\times 2$ game \cite{jafari2001no}. Thus, when both agents follow algorithm $\mathcal{A}$, the strategies converge to the pure Nash equilibrium, i.e., $(2,2)$ entry leading to the Nash outcome of $(5,5)$. Thus, $U_1(\mathcal{A},\mathcal{A})=5$.

    Now consider Algorithm $\mathcal{A}'$ which, at each time $t$, has agent agent $1$ play row $j=1$. We now show that, for algorithm $\mathcal{A}$ and a given constant $c$, equation \eqref{eq:secure_algo} does not hold.

    Suppose that at time $t$, agent $2$ selects action $k=1$. Given game $\mathcal{G}$ in Figure \ref{fig:RM_fragile}, the instantaneous regret for agent $2$ for not selecting column $k=2$ is
    \begin{align*}
        \delta_2^t(2) = r^2_{1,2} - r^2_{1,1} = -1.
    \end{align*}
    Similarly, at time $t$, if agent $2$ selects action $k=2$, then the instantaneous regret for agent $2$ for not selecting column $k=1$ is
    \begin{align*}
        \delta_2^t(1) = r^2_{1,1} - r^2_{1,2} = 1.
    \end{align*}
    Since the action of agent $1$ does not change at any time $t$, it follows that at every time $t$ at which agent $2$  selected column $2$, agent $2$ experienced a positive regret. Thus, as $t\to \infty$, from equation \eqref{eq:rm_prob}, $p_{t+1}^2(1)\to 1$. This implies that $U_1(\mathcal{A}',\mathcal{A})=r^1_{1,1}$. Since $r^1_{1,1}=5(c+1)$ it follows that
    \begin{align*}
        \frac{U_1(\mathcal{A}',\mathcal{A})}{U_1(\mathcal{A},\mathcal{A})} = c+1\implies \sup_{\mathcal{G},\mathcal{A}'} \frac{U_1(\mathcal{A}',\mathcal{A})}{U_1(\mathcal{A},\mathcal{A})} \geq c+1.
    \end{align*}
    This means that the condition in equation \eqref{eq:secure_algo} can never hold for any given constant $c$ and the result is established.
\end{proof}

\section{Proof of Lemma \ref{lem:punishment}}
\begin{proof}
    Without loss of generality, suppose that agent $1$ deviates from algorithm $\mathcal{A}$.
    First, consider that algorithm $\mathcal{A}$ enters the punishment phase from the exploitation sub-phase. Then, agent $2$ selects action according to the minimax strategy, defined in Definition \ref{def:minimax}, on matrix $\hat{R}^1=R^1$. As $t\to\infty$, $U_1(\mathcal{A}',\mathcal{A})\to \bar{V}_1(\hat{R}^1)$. Since $\bar{V}_1^p(\hat{R}^1)\geq \bar{V}_1(\hat{R}^1)$ \cite{hespanha2017noncooperative} and given the condition in equation \eqref{eq:minimax_pure}, we obtain 
    \begin{equation}\label{eq:bound_5.2}
        \frac{U_1(\mathcal{A}',\mathcal{A})}{U_1(\mathcal{A},\mathcal{A})}= \frac{\bar{V}_1(R^1)}{U_1(\mathcal{A},\mathcal{A})}\leq \frac{\bar{V}_1^p(R^1)}{U_1(\mathcal{A},\mathcal{A})}\leq 1.
    \end{equation}
    
    Note that equation \eqref{eq:bound_5.2} holds even in the case when algorithm $\mathcal{A}$ enters the punishment phase from the exploration sub-phase and there exists a time $t$ at which the matrix $R^1$ is completely known by agent $2$. 
    Further note that, in the case when none of the rows of the payoff matrix $R^1$ is completely known by agent $2$, at any time $t$, there is a positive probability that a new entry of $R^1$ will be known. This implies that there exists a time $t$ at which at least one of the row of matrix $R^1$ will be completely known by agent $2$.
    Thus, in what follows, we will consider the case for which the following jointly holds:
    \begin{itemize}
        \item \textbf{P1:} algorithm $\mathcal{A}$ enters the punishment phase from the exploration sub-phase,
        \item \textbf{P2:} at least one of the rows of $R^1$ is completely known by agent $2$, and
        \item \textbf{P3:} the matrix $R^1$ is not completely known by agent $2$ at any time $t$.
    \end{itemize}
    Without loss of generality, let the $j$th row of matrix $R^1$ be completely known to agent $2$. 
    Given \textbf{P3}, let $\tau$ denote the time when an entry of matrix $R^1$ was revealed and no other entry of matrix $R^1$ is revealed for any time $t>\tau$. Further, as $t\to\infty$ and since no new entry of $R^1$ is revealed after time $\tau$, $U_1(\mathcal{A}',\mathcal{A})\to \bar{V}_1(\tilde{R}^1)$.

    Since $\bar{V}_1^p(\tilde{R}^1)\geq \bar{V}_1(\tilde{R}^1)$ \cite{hespanha2017noncooperative}, we now show that $\bar{V}_1^p(\tilde{R}^1)\leq \bar{V}_1^p(R^1)$. Suppose that entry $(j',k)$ for any $j'\neq j$ of matrix $R^1$ is not known by agent $2$. Let $\tilde{R}^1_{j'}$ denote the matrix if the entry $(j',k)$ was known by agent $2$. Then, if $r^1_{j',k} > r^1_{j,k}$, it follows that $\bar{V}_1^p(\tilde{R}^1_{j'})> \bar{V}_1^p(\tilde{R}^1)$. This is because the entry $r^1_{j',k}=0$ in matrix $\tilde{R}^1$. Further, if $r^1_{j',k} \leq r^1_{j,k}$, it follows that $\bar{V}_1^p(\tilde{R}^1_{j'}) = \bar{V}_1^p(\tilde{R}^1)$. Thus, for any $(j',k)$ entry that is not known by agent $2$, we have $\bar{V}_1^p(\tilde{R}^1_{j'})\geq \bar{V}_1^p(\tilde{R}^1)$ which further implies that $\bar{V}_1^p(R^1)\geq \bar{V}_1^p(\tilde{R}^1)$. Given the condition that $\bar{V}_1^p(R^1)\leq U_1(\mathcal{A},\mathcal{A})$, we obtain 
    \begin{align*}
        \frac{\bar{V}_1(\tilde{R}^1)}{U_1(\mathcal{A},\mathcal{A})}\leq \frac{\bar{V}_1^p(\tilde{R}^1)}{U_1(\mathcal{A},\mathcal{A})}\leq \frac{\bar{V}_1^p(R^1)}{U_1(\mathcal{A},\mathcal{A})}\leq 1.
    \end{align*}
    This concludes the proof.
\end{proof}

\section{Proof of Theorem \ref{thm:Sec_FM}}
\begin{proof}
    Without loss of generality, suppose agent $1$ deviates from Algorithm R-GFP. From the definition of Algorithm R-GFP and given the perfect monitoring setting, any deviation is ensured to be detected. Thus, it is ensured that the algorithm enters the punishment phase once agent $1$ deviates. Thus, from Lemma \ref{lem:punishment}, Algorithm R-GFP is perfectly rational.

    If none of the agents deviate, after time $t=|A_1||A_2|$, Algorithm R-GFP enters exploitation phase. In the exploitation phase, the agents select their actions according to fictitious play for all time $t>|A_1||A_2|$. This implies that the corresponding strategies of the agents following Algorithm R-GFP will be the same as when the agents follow fictitious play and the result follows.
\end{proof}

\section{Proof of Theorem \ref{thm:Sec_RM}}
The following result will be useful in the proof.

\begin{lemma}[Dvoretzky–Kiefer–Wolfowitz inequality\cite{kosorok2008introduction}]
    Given a natural number $n$, let $X_1,X_2,\dots,X_n$ be independent and identically distributed random variables with CDF $\mathcal{F}$. Let $F_n$ denote the associated empirical distribution function defined by
    \begin{align*}
        F_n(x) = \frac{1}{n}\sum_{i=1}^n\textbf{1}_{\{X_i\leq x\}}. 
    \end{align*}
    Then, 
    \begin{align*}
        Pr\left(\sup_{x\in \mathbb{R}} |\mathcal{F}(x)-F_n(x)|>\epsilon \right)\leq 2e^{-2n\epsilon^2}, \forall \epsilon>0.
    \end{align*}
    \label{lem:DKW_inequality}
\end{lemma}

We now provide the proof of Theorem \ref{thm:Sec_RM}.

\begin{proof}
    Without loss of generality, suppose that agent $1$ deviates from Algorithm R-RM. 
    First, if agent $1$ deviates during exploration sub-phase, then any deviation is guaranteed to be detected. This means that, if agent $1$ deviates during exploration sub-phase, then Algorithm R-RM is ensured to enter the punishment phase. Thus, it follows from Lemma \ref{lem:punishment} that, in this case, Algorithm R-RM is perfectly rational. We now show that Algorithm R-RM is perfectly rational when agent $1$ deviates during exploitation sub-phase.

    Observe that in any epoch $t$, in the worst-case, agent $1$ can select its actions such that the deviation between the empirical CDF and the actual CDF is within $\epsilon_t$. Mathematically, in any epoch $t$ and in the worst-case, agent $1$ can selects its action such that at after $N_t$ iterations, the following condition does not hold:
    \begin{align}
        \sup_{x\in \mathbb{R}} |\mathcal{F}_t^1(x)-F_t^1(x)| > \epsilon_t.
    \end{align}
    For any epoch $t$, as $\epsilon_t=\tfrac{1}{t}$, it follows that as $t\to \infty$, $\epsilon_t\to 0$. Thus, for a high value of $t$, there are two cases.

    \textbf{Case 1:} The first case is that after some epoch $t$, since $\epsilon_t\approx 0$, the condition defined in equation \eqref{eq:CDF_comp} holds in which case Algorithm R-RM enters punishment phase. In this case, from Lemma \ref{lem:punishment}, Algorithm R-RM is perfectly rational.

    \textbf{Case 2:} The second case is that, since $\epsilon_t\to 0$ as $t\to\infty$, agent $1$ starts selecting actions according to regret matching and does not deviate. This means that from this moment on, both agents select their actions according to regret matching algorithm. We now show that in this case, the strategies obtained from Algorithm R-RM converge to $\pi_1^*$ and $\pi_2^*$, with probability $1-\delta$. 
    
    Intuitively, the idea is to show that the probability with which Algorithm R-RM enters punishment phase when the agents do not deviate is very small.
    For any epoch $t$, using Lemma \ref{lem:DKW_inequality}, the probability that condition defined in equation \eqref{eq:CDF_comp} holds is
    \begin{align*}
        Pr\left(\sup_{x\in \mathbb{R}} |\mathcal{F}(x)-F_{N_t}(x)|>\epsilon _t\right)\leq 2e^{-2N_t\epsilon_t^2}.
    \end{align*}
    Then, by taking the union bound over all possible epochs $t$, it follows that the probability that the condition defined in equation \eqref{eq:CDF_comp} never holds is at least $1-t2e^{-2N_t\epsilon_t^2}$. For some positive real constants $c_1$ and $c_2$, by selecting $N_t$ as
    \begin{align*}
        N_t = \frac{c_1\log(\frac{c_2t}{\delta})}{\epsilon^2_t},
    \end{align*}
    we obtain that the probability that the condition defined in equation \eqref{eq:CDF_comp} never holds is at least $1-\frac{2\delta}{c_2t^{2c_1-1}}$. By selecting $c_1$ and $c_2$ such that $\frac{2\delta}{c_2t^{2c_1-1}}\geq \delta$, it follows that with probability at least $1-\delta$, the condition defined in equation \eqref{eq:CDF_comp} never holds until epoch $t$. This means that, with probability $1-\delta$, Algorithm R-RM does not enter the punishment phase until epoch $t$. Thus, if $\pi_1^*$ and $\pi_2^*$ converge to an equilibrium, the strategies obtained from Algorithm R-RM will also converge to the equilibrium. 

    Finally, the proof for the case when no agent deviates and the strategies obtained from Algorithm R-RM converges to the equilibrium, with probability $1-\delta$, if $\pi_1^*$ and $\pi_2^*$ converge to an equilibrium is analogous to that of Case 2. This concludes the proof.
\end{proof}

\section{Proof of Theorem \ref{thm:imperfect_games}}
\begin{figure}[t]
\centering
  \begin{subfigure}[b]{0.3\textwidth}
  \centering
    \includegraphics[width=\linewidth]{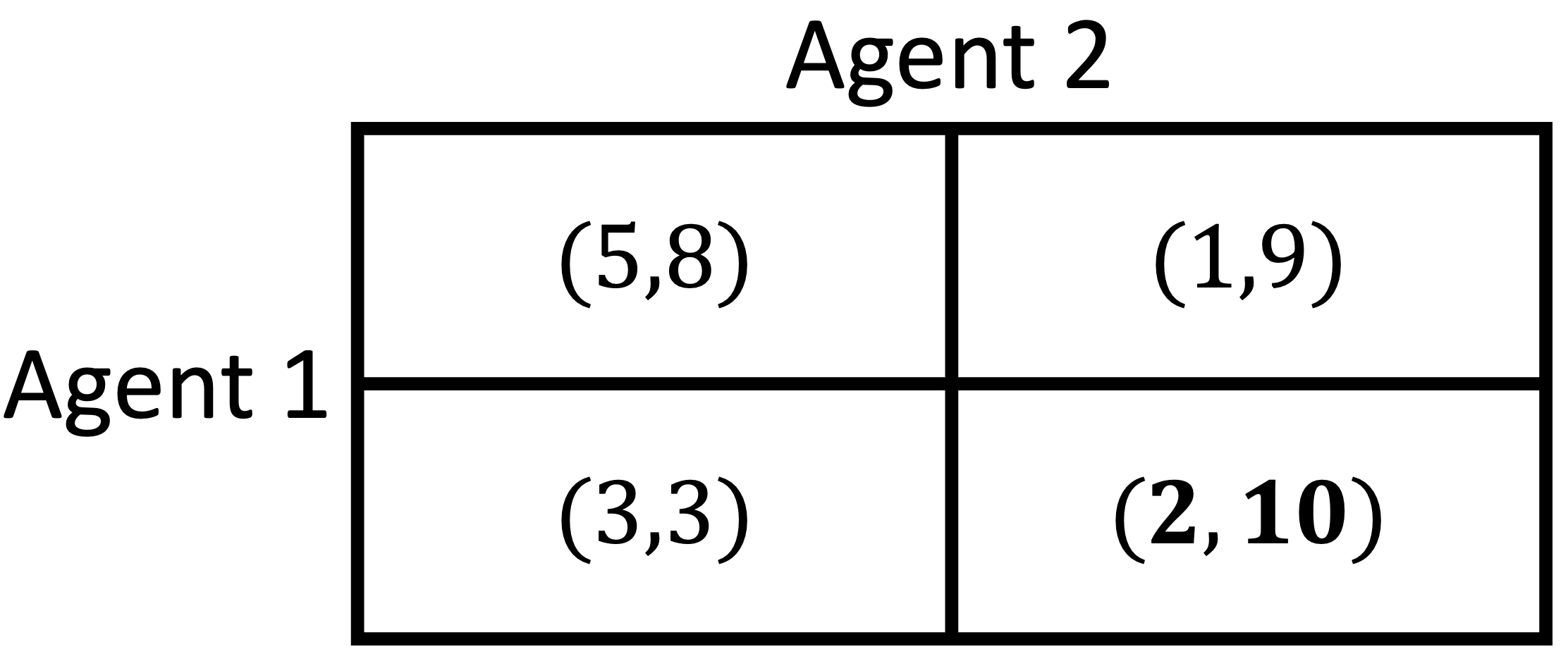}
    \caption{Game $\mathcal{G}_1$.}
  \end{subfigure}
  \begin{subfigure}[b]{0.3\textwidth}
    \includegraphics[width=\linewidth]{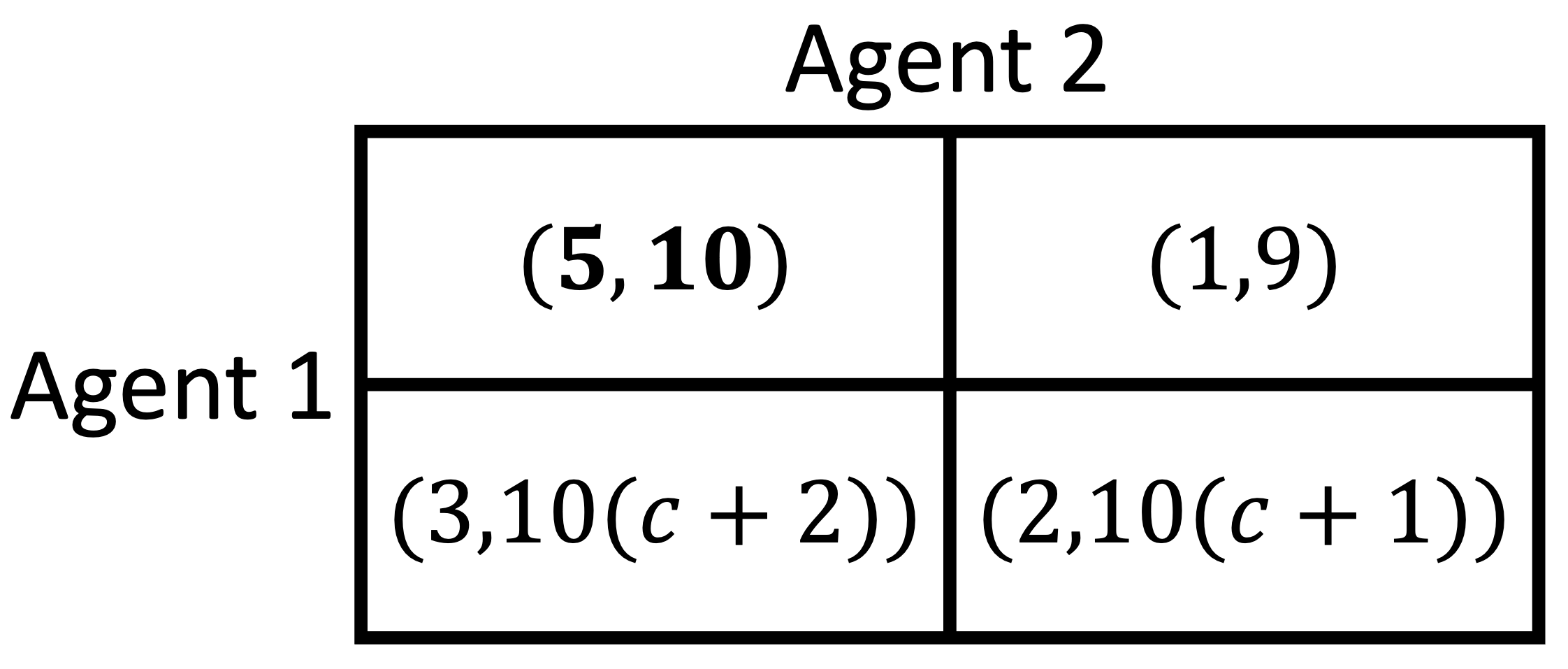}
    \caption{Game $\mathcal{G}_2$.}
  \end{subfigure}
  \caption{Games $\mathcal{G}_1$ and $\mathcal{G}_2$ for the proof of Theorem \ref{thm:imperfect_games}. The equilibrium entries are highlighted in bold.}
  \label{fig:imperfect_games}
\end{figure}
\begin{proof}
    Without loss of generality, suppose that agent $2$ deviates from an algorithm $\mathcal{A}$ and follows algorithm $\mathcal{A}'$.
    Consider two games $\mathcal{G}_1$ and $\mathcal{G}_2$ as depicted in Figure \ref{fig:imperfect_games}. In game $\mathcal{G}_1$ (resp. $\mathcal{G}_2$), the entry $(2,2)$ (resp. entry $(1,1)$) is the only possible equilibrium (because of domination) and suppose that, if both agents select their actions according to algorithm $\mathcal{A}$, their respective strategies converge to the entries corresponding to the equilibrium of the game.

    Suppose agent $2$ always selects column $2$ and the game is $\mathcal{G}_2$. Further, even by assuming that Algorithm $\mathcal{A}$ has the information that the game selected is either $\mathcal{G}_1$ and $\mathcal{G}_2$ and completely knows its own payoff matrices, $\mathcal{A}$ cannot determine the actual game being played by the agents. This is because of the imperfect monitoring setting and that the payoffs for agent $1$ is identical in both $\mathcal{G}_1$ and $\mathcal{G}_2$. Thus, given that agent $1$ can only observe agent $2$'s actions and since agent $2$ selects only column $2$, algorithm $\mathcal{A}$ selects row $2$ for agent $1$. This is because, given that agent $2$ selects column $2$, selecting row $1$ yields a lower payoff for agent $1$.
    This means that $U_2(\mathcal{A},\mathcal{A}')=10(c+1)$. Thus, even by restricting the set of games to only $\mathcal{G}_1$ and $\mathcal{G}_2$, it follows that $\sup_{\mathcal{G}_1,\mathcal{G}_2,\mathcal{A}'} \frac{U_{2}(\mathcal{A},\mathcal{A}')}{U_{2}(\mathcal{A},\mathcal{A})} = c+1$. This implies that $\sup_{\mathcal{G},\mathcal{A}'} \frac{U_{2}(\mathcal{A},\mathcal{A}')}{U_{2}(\mathcal{A},\mathcal{A})} > c$, for any constant $c$ meaning that $\mathcal{A}$ is not $c$-rational for any constant $c$. This concludes the proof.
\end{proof}

\section{Additional Numerical Results}
In this section, we provide additional numerical results to illustrate the analytical results established in this work.

\begin{table}[t]
\caption{Numerical results for Algorithm R-GFP. Fictitious play is denoted as FP.}
\label{tb:numerics_RFP}
\begin{center}
\begin{small}
\begin{sc}
\begin{tabular}{lccccr}
\toprule
Game & $U_1(\mathcal{A}',\text{FP})$ & $U_1(\text{R-GFP},\text{R-GFP})$ & $U_1(\mathcal{A}',\text{R-GFP})$ & $U_1(\mathcal{A}'_{\text{explore}},\text{R-GFP})$ \\
\midrule
$\mathcal{G}_1^{\text{GFP}}$    & $99.3$ & $25.3$ & $1.2$ & $1.1$ \\
$\mathcal{G}_2^{\text{GFP}}$ &   $42$ & 34& $27.5$ & $27.5$\\
$\mathcal{G}_3^{\text{GFP}}$    & $72$ & 66& $41.4$ & $10.4$ \\
\bottomrule
\end{tabular}
\end{sc}
\end{small}
\end{center}
\end{table}

\begin{figure}[t]
\centering
  \begin{subfigure}[b]{0.29\columnwidth}
  \centering
    \includegraphics[width=\linewidth]{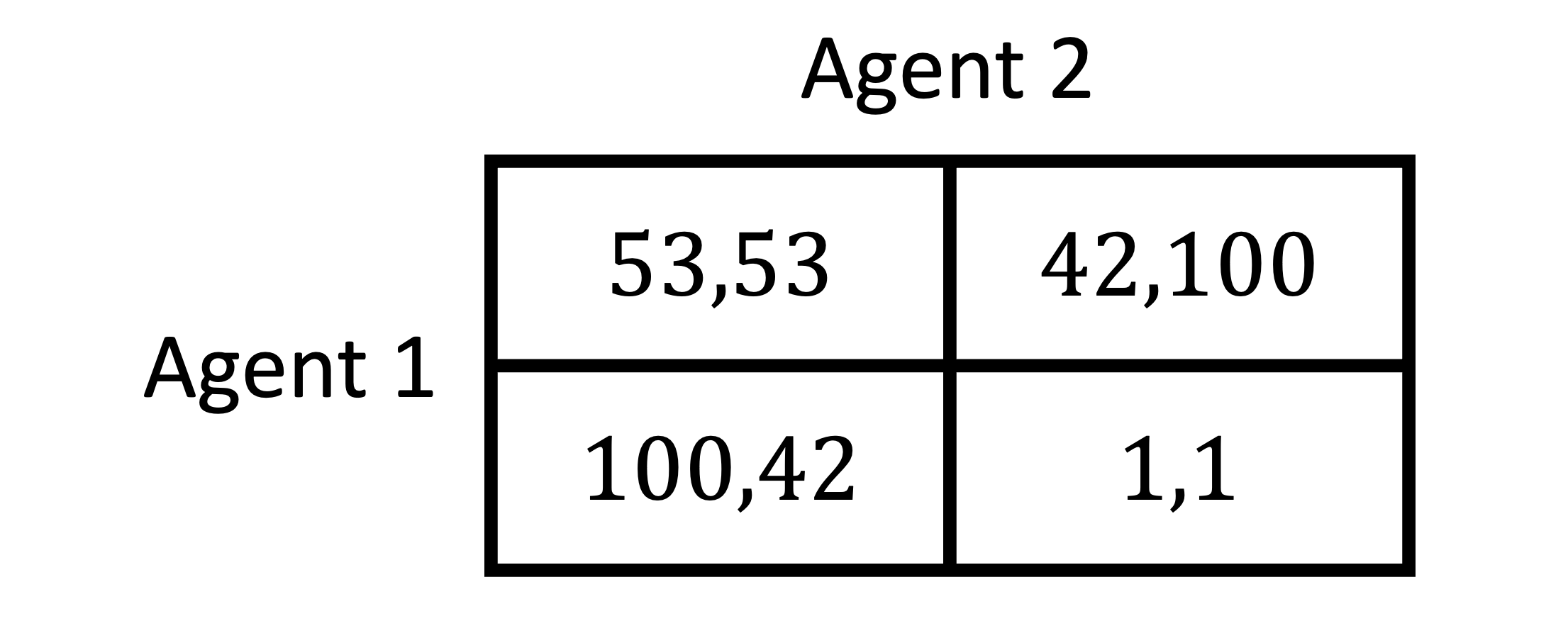}
    \caption{Game $\mathcal{G}_1^{\text{GFP}}$.}
  \end{subfigure}
  \hfill 
  \begin{subfigure}[b]{0.33\columnwidth}
  \centering
    \includegraphics[width=\linewidth]{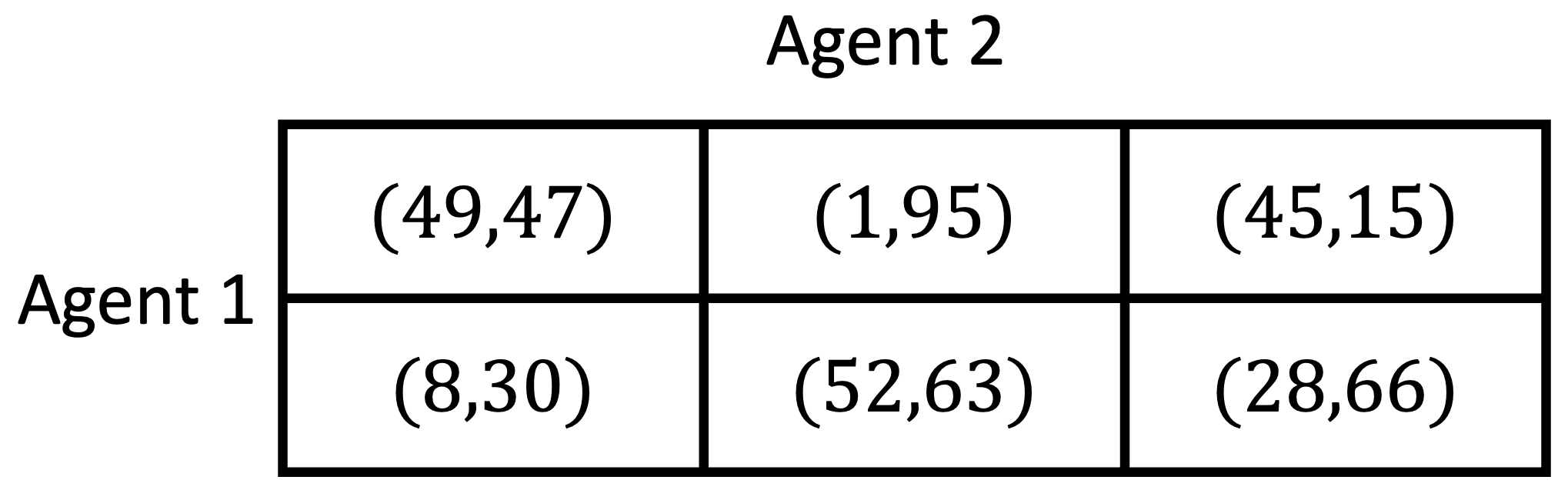}
    \caption{Game $\mathcal{G}_2^{\text{GFP}}$.}
  \end{subfigure}
  \hfill 
  \begin{subfigure}[b]{0.33\columnwidth}
  \centering
    \includegraphics[width=\linewidth]{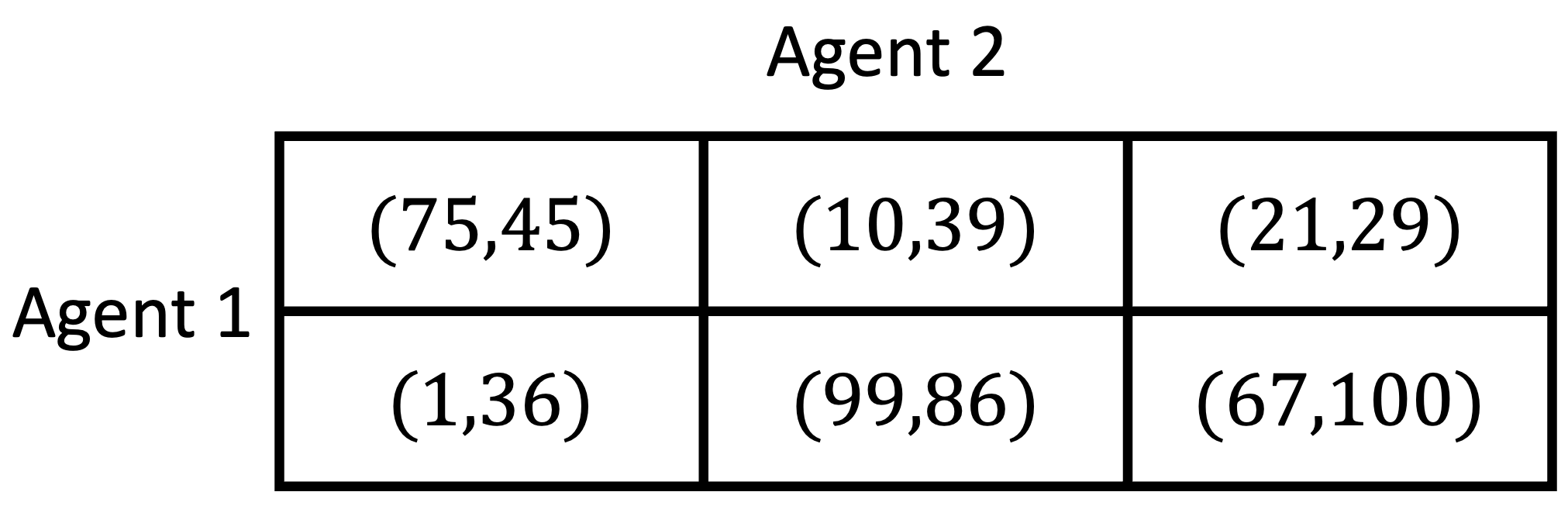}
    \caption{Game $\mathcal{G}_3^{\text{GFP}}$.}
  \end{subfigure}
  \caption{Games $\mathcal{G}_1^{\text{GFP}}$, $\mathcal{G}_2^{\text{GFP}}$, and $\mathcal{G}_3^{\text{GFP}}$ for Table \ref{tb:numerics_RFP}.}
  \label{fig:Games_FP_numerics}
\end{figure}

\begin{figure}[t]
\centering
  \begin{subfigure}[b]{0.29\columnwidth}
  \centering
    \includegraphics[width=\linewidth]{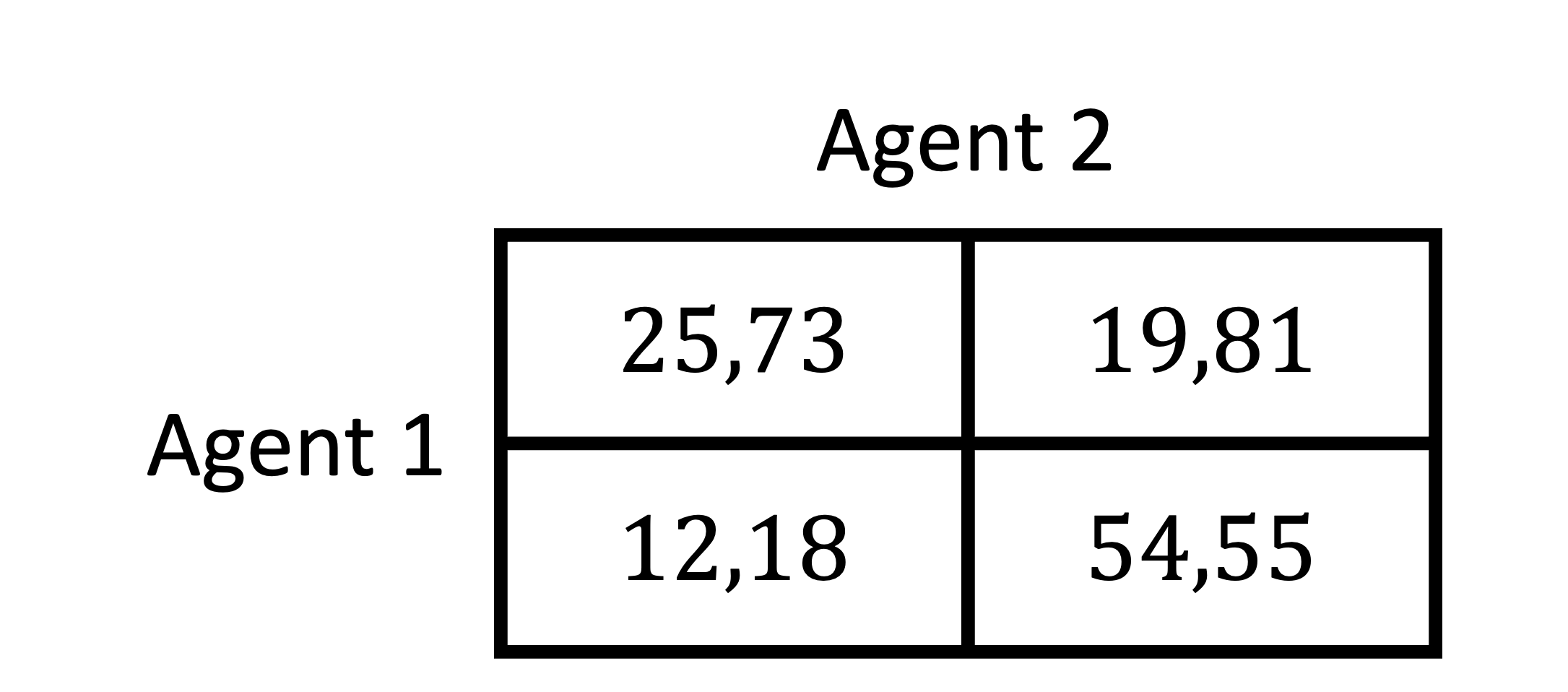}
    \caption{Game $\mathcal{G}_1^{\text{RM}}$.}
  \end{subfigure}
  \hfill 
  \begin{subfigure}[b]{0.33\columnwidth}
  \centering
    \includegraphics[width=\linewidth]{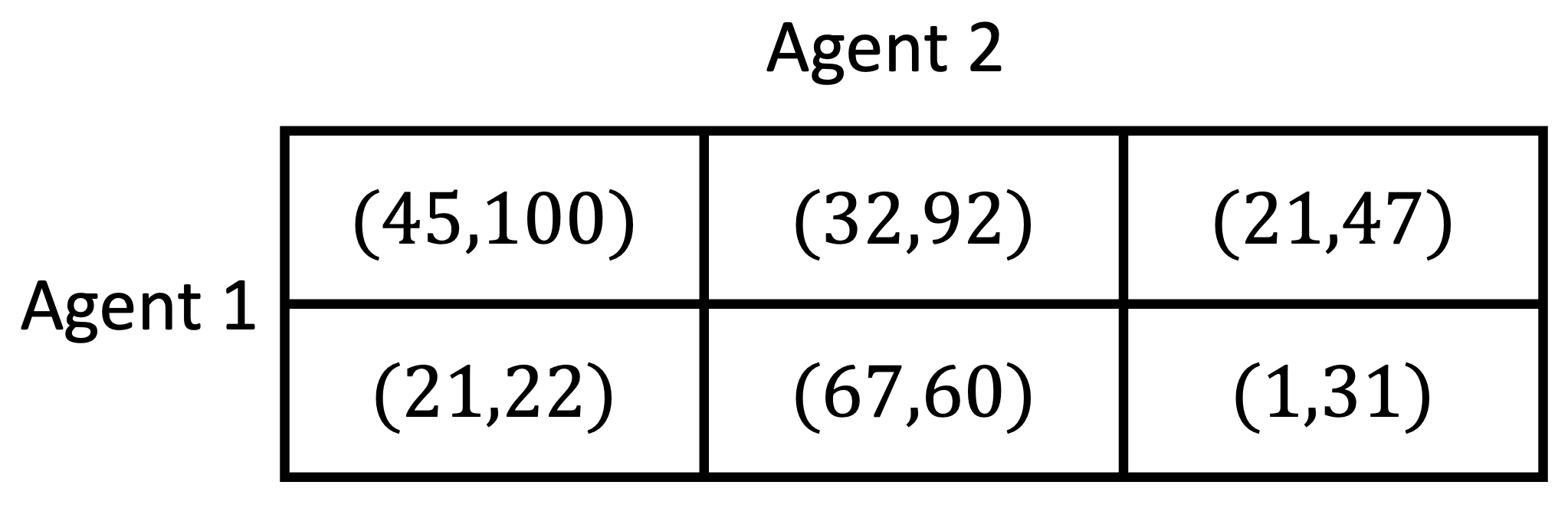}
    \caption{Game $\mathcal{G}_2^{\text{RM}}$.}
  \end{subfigure}
  \hfill 
  \begin{subfigure}[b]{0.33\columnwidth}
  \centering
    \includegraphics[width=\linewidth]{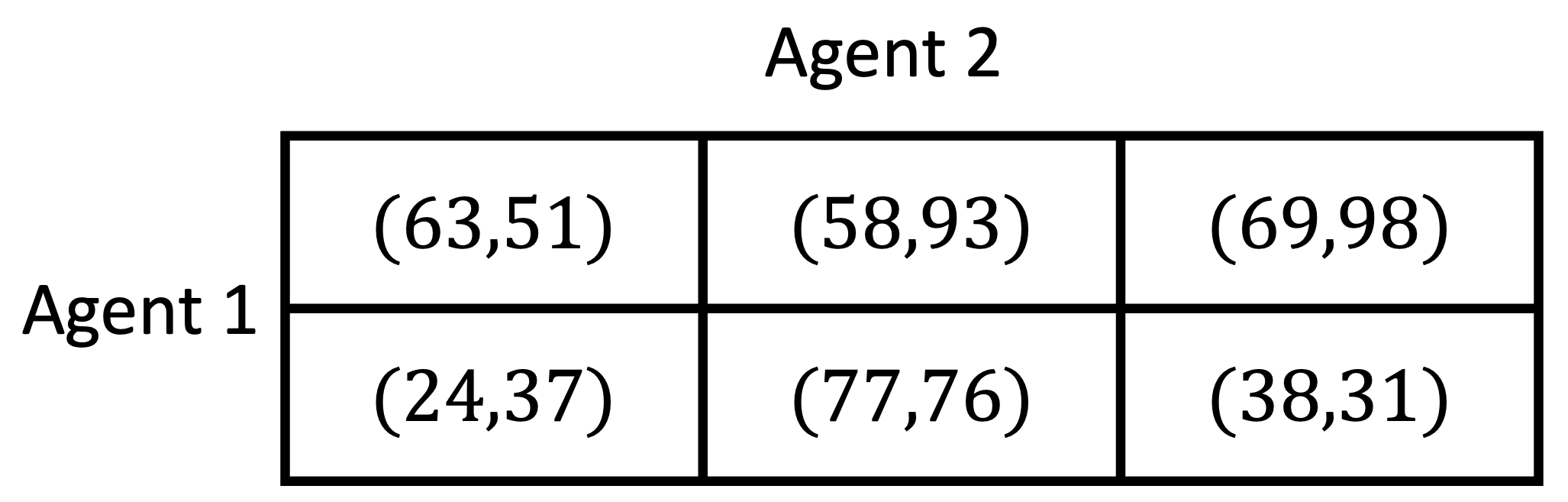}
    \caption{Game $\mathcal{G}_3^{\text{RM}}$.}
  \end{subfigure}
  \caption{Games $\mathcal{G}_1^{\text{RM}}$, $\mathcal{G}_2^{\text{RM}}$, and $\mathcal{G}_3^{\text{RM}}$ for Table \ref{tb:numerics_RRM}.}
  \label{fig:Games_RM_numerics}
\end{figure}

Table \ref{tb:numerics_RFP} and Table \ref{tb:numerics_RRM} illustrates the numerical results for Algorithm R-GFP and Algorithm R-RM, respectively. The games considered in Table \ref{tb:numerics_RFP} and Table \ref{tb:numerics_RRM} were generated using GAMUT \cite{nudelman2004run} and are illustrated in Figure \ref{fig:Games_FP_numerics} and Figure \ref{fig:Games_RM_numerics}, respectively.

Table \ref{tb:numerics_RFP} presents different values of payoffs obtained by agent $1$ (adversary) for different cases. 
We consider that the agents use the entire history and thus, the generalized fictitious play is equivalent to fictitious play and implement the strategy for agent $1$ as described in \cite{vundurthy2023intelligent}. In game $\mathcal{G}_1^{\text{GFP}}$ and $\mathcal{G}_2^{\text{GFP}}$, the payoff obtained by the adversary upon deviating during the exploration sub-phase is approximately equal to when it deviates during the exploitation sub-phase. This is because the matrix $R^1$ is eventually completely known by agent $2$ as $t\to \infty$. In game $\mathcal{G}_3^{\text{GFP}}$, the adversary never selects row $2$ meaning that the matrix $R^1$ is never fully known to agent $2$. Thus, for game $\mathcal{G}_3^{\text{GFP}}$, the payoff obtained by the adversary upon deviating during the exploration sub-phase is less than when it deviates during exploitation sub-phase. Finally, for all games, $U_1(\mathcal{A}',\text{R-GFP})<U_1(\text{R-GFP},\text{R-GFP})$ implying that Algorithm R-GFP is perfectly rational.

\begin{table}[h]
\caption{Numerical results for Algorithm R-RM. Regret-matching is denoted as RM.}
\label{tb:numerics_RRM}
\begin{center}
\begin{small}
\begin{sc}
\begin{tabular}{lccccr}
\toprule
Game & $U_2(\text{RM},\mathcal{A}')$ & $U_2(\text{R-RM},\text{R-RM})$ & $U_2(\text{R-RM},\mathcal{A}')$ & $U_2(\text{R-RM},\mathcal{A}'_{\text{explore}})$ \\
\midrule
$\mathcal{G}_1^{\text{RM}}$    & $72$ & 54.9& $18.2$ & $18$ \\
$\mathcal{G}_2^{\text{RM}}$    & $98.5$ & $60.1$ & $22.2$ & $22.1$   \\
$\mathcal{G}_3^{\text{RM}}$ &   $97.8$ & $79$ & $31.5$ & $30.9$ \\
\bottomrule
\end{tabular}
\end{sc}
\end{small}
\end{center}
\end{table}

Table \ref{tb:numerics_RRM} presents different values of payoffs obtained by agent $2$ (adversary) for various cases. 
Upon deviation, agent $2$ follows the strategy as described in equation \eqref{eq:strat_adv_RM}. Similar to Algorithm R-GFP, for all games in Table \ref{tb:numerics_RRM}, $U_2(\text{R-RM},\mathcal{A}')<U_2(\text{R-RM},\text{R-RM})$ implying that Algorithm R-RM is perfectly rational. Finally, for all games in Table \ref{tb:numerics_RRM}, the value of agent $2$ when it deviates during exploration sub-phase is approximately equal to the vale when agent $2$ deviates during exploitation sub-phase as matrix $R^2$. This is because the strategy followed by agent $2$ does not change in both exploration and exploitation sub-phase once agent $1$ follows minimax strategy.



\end{document}